\documentclass[prd,onecolumn,floatfix,superscriptaddress]{revtex4}

\usepackage{graphicx}
\usepackage{epsfig}
\usepackage{bm}
\usepackage{amssymb}
\usepackage{float}
\usepackage{amsmath}
\usepackage{dcolumn}
\usepackage{cancel}
\usepackage[colorlinks]{hyperref}
\usepackage{color}

\begin{document}

\title{Stability of motion and thermodynamics in charged  black holes 
in  $f(T)$ gravity}

\author{G.G.L. Nashed}
\email{nashed@bue.edu.eg}
\affiliation{Centre for Theoretical Physics, The British University, P.O. Box
43, El Sherouk City, Cairo 11837, Egypt}
\author{Emmanuel N. Saridakis}
\email{msaridak@noa.gr}
\affiliation{National Observatory of Athens, Lofos Nymfon, 11852 Athens,
	Greece}
\affiliation{CAS Key Laboratory for Researches in Galaxies and Cosmology,
Department of Astronomy, University of Science and Technology of China, Hefei,
Anhui 230026, P.R. China}
\affiliation{School of Astronomy, School of Physical Sciences,
University of Science and Technology of China, Hefei 230026, P.R. China}

\begin{abstract}

We investigate the stability of motion   and  the thermodynamics in the 
case of spherically symmetric
solutions in $f(T)$ gravity  using the perturbative approach. We
consider small deviations from   general relativity
and we extract charged black hole solutions for two charge profiles, namely
with or without a perturbative correction in the charge distribution.
We examine their asymptotic behavior, we extract various torsional and
curvature invariants, and we calculate the energy and the mass of the
solutions.  Furthermore, we study the   stability  of motion around the 
obtained  solutions, by analyzing  the geodesic
deviation, and we extract the unstable regimes in
the parameter space. We calculate the inner
(Cauchy) and outer (event) horizons, showing that for larger  deviations from
general relativity  or larger charges, the horizon disappears and the central
singularity
becomes a naked one. Additionally, we perform a detailed thermodynamic
analysis examining   the temperature, entropy, heat capacity and Gibb's free
energy.   Concerning  the heat capacity  we find that for larger deviations from
general relativity it is always positive,    and this
shows that $f(T)$ modifications improve the thermodynamic     stability, which 
is
 not the case in other classes of modified gravity.


\end{abstract}
\maketitle

\section{Introduction}

There are both theoretical and observational motivations for the construction
of gravitational modifications, namely of extended theories of gravity that
possess general relativity as a particular limit, but which in general exhibit
a richer structure \cite{CANTATA:2021ktz}. The first is based on the fact that
since general relativity is non-renormalizable one could hope that more
complicated extensions of it would improve the renormalizability properties
\cite{Stelle:1976gc,Addazi:2021xuf}. The second motivation is related to the 
observed
features of the Universe, and in particular the need to describe its two
accelerated phases, namely one at early times (inflation) and one at late times
(dark energy era).  The usual approach in  the construction of  gravitational
modifications is to start from the Einstein-Hilbert action and  extend it in
various ways   \cite{Capozziello:2011et}. Nevertheless, one can
  start from the equivalent torsional formulation of gravity, and in particular
from the Teleparallel Equivalent of General Relativity (TEGR)
\cite{Unzicker:2005in, Aldrovandi:2013wha,Shirafuji:1997wy, Maluf:2013gaa}, and modify it
accordingly, obtaining $f(T)$ gravity \cite{Cai:2015emx,
Bengochea:2008gz, Linder:2010py},   $f(T,T_G)$ gravity \cite{Kofinas:2014owa},
$f(T,B)$ gravity \cite{Bahamonde:2015zma, Karpathopoulos:2017arc,
Boehmer:2021aji}, scalar-torsion theories \cite{Geng:2011aj, Hohmann:2018rwf,
Bahamonde:2019shr}, etc. Torsional gravity, can lead to interesting cosmological
phenomenology and hence it has attracted a large amount of research
\cite{Cai:2015emx,Zheng:2010am,ElHanafy:2015egm,
Bamba:2010wb, Capozziello:2011hj,Wei:2011aa, Amoros:2013nxa,
Otalora:2013dsa, Bamba:2013jqa, Li:2013xea,
Malekjani:2016mtm, Farrugia:2016qqe,Nashed:2014sea,Saridakis:2017eqb, Qi:2017xzl,Cai:2018rzd,
Abedi:2018lkr, El-Zant:2018bsc, Anagnostopoulos:2019miu,
Cai:2019bdh, Yan:2019gbw,Awad:2017ign,ElHanafy:2019zhr, Wang:2020zfv,
ElHanafy:2020pek,Nashed:2009hn,Hashim:2020sez, Ren:2021tfi}.

Additionally, torsional and $f(T)$ gravity exhibit
  novel and interesting
black hole and spherically symmetric solutions too \cite{Boehmer:2011gw,
Gonzalez:2011dr, Capozziello:2012zj,Ferraro:2011ks, Wang:2011xf,
Atazadeh:2012am,Nashed:2005kn,
Rodrigues:2013ifa, Nashed:2013bfa, Nashed:2014iua, Junior:2015fya,
Kofinas:2015hla, Das:2015gwa,Awad:2017sau, Nashed:2018efg,Rani:2016gnl, Rodrigues:2016uor, Mai:2017riq,
Singh:2019ykp, Nashed:2020kjh, Bhatti:2018fsc, Ashraf:2020yyo,
Ditta:2021wfl,Ren:2021uqb}.
In particular,   spherically symmetric solution with a constant torsion
scalar $T$  have been studied in
\cite{2011PhRvD..84h3518F,Nashed:uja,2013PhRvD..88j4034N}, while cylindrically
charged black holes solutions using quadratic and cubic forms of $f(T)$ have
been derived \cite{Nashed:2019zmy,Nashed:2018cth,Awad:2017tyz,Nashed:2016tbj}.
Moreover,   by using the Noether's symmetry   approach, static spherically black
hole solutions have been investigated in \cite{Paliathanasis:2014iva}.
 In similar lines, the research  of static spherically symmetric solutions
using   $f(T)$ corrections on TEGR was the focus of interest in many studies
using the  perturbative approach
\cite{DeBenedictis:2016aze,Bahamonde:2020bbc,Bahamonde:2019zea,
 2015PhRvD..91j4014R, Nashed:2021rgn}, while vacuum
regular BTZ  black hole solutions in Born-Infeld gravity have been extracted in
\cite{Bohmer:2019vff,Bohmer:2020jrh}.

Although       spherically symmetric solutions in $f(T)$ gravity
has been investigated in many works,  the stability of motion around 
them has not been examined in
detail. This issue is quite crucial, having in mind that modifications of
general relativity are known to  present various instabilities   is various
regimes of the parameter space. Hence, in this work we aim to derive charged
spherically symmetric
solution  in $f(T)$ gravity using the perturbative approach, and then examine
the   stability of motion and thermodynamic properties.

The arrangement of the manuscript is as follows: in Section \ref{S2}  we
extract the charged black-hole solutions for $f(T)$ gravity, using the
perturbative approach, for two charge profiles.
  In  Section
 \ref{S4}  we study the properties of the extracted perturbative
solutions, and in particular their asymptotic forms, the invariants, and their
energy.  In Section \ref{S6336b} we proceed to the investigation of
the  stability of motion around   the
solutions, by extracting and analyzing the geodesic deviation. Moreover, in
Section  \ref{S5} we study in detail the thermodynamic properties, focusing on
the temperature,   entropy,   heat capacity and   Gibbs free energy.
  The final Section \ref{Conclusions} is reserved for
  conclusions and discussion.

\section{Charged black hole solutions in  $f(T)$ gravity}\label{S2}

Let us extract charged black hole solutions following the perturbative
approach. As usual, in torsional gravity as the dynamical field we use the
orthonormal tetrad, whose components   are     $h_a{}^\mu$, with  Latin
indices     (from 0 to 3) denoting  the tangent space and
  Greek indices (from 0 to 3)   marking the coordinates
on the manifold.  The relation between the tetrad and the manifold metric is
  $g_{\alpha \beta}=\eta_{i j} h^i{}_\alpha h^j{}_\beta$, with
$\eta_{i j}$  the Minkowski metric $\eta_{i
j}=diag.(-1,+1,+1,+1)$. The torsion tensor is given as
$
	T^a{}_{\mu\nu}:= \partial_{\mu}h^a{}_{\nu}-\partial_{\nu}h^a{}_{\mu}$.

The action of   $f(T)$ gravity, alongside a minimally coupled elecromagnetic
sector,  is \cite{Gonzalez:2011dr,Capozziello:2012zj}
\begin{align}\label{eq:fT1}
	S_{f(T)} = \int d^4x\ |h| \left( \frac{1}{2\kappa^2}  f(T) + {\cal
F}\right)\,,
\end{align}
with $\kappa=8\pi G$ the gravitational constant, and $|h|=det
(h^a{}_\mu)=\sqrt{-g}$. The   torsion scalar $T$
is written as
$	T =  T^a{}_{\mu\nu}S_a{}^{\mu\nu} $
in terms of the superpotential
$S_a{}^{\mu\nu} = \frac{1}{2}(K^{\mu\nu}{}_a - h_a{}^\mu
T_\lambda{}^{\lambda\nu} + h_a{}^\nu T_\lambda{}^{\lambda\mu})$,
with the  contortion tensor  being
$
{K^{\mu \nu}}_\alpha:=
-\frac{1}{2}\left({T^{\mu \nu}}_\alpha-{T^{\nu
\mu}}_\alpha-{T_\alpha}^{\mu \nu}\right)$.
Additionally,   ${\cal {\cal F}}$ is  the gauge-invariant
Lagrangian of electromagnetism given as ${\cal F} =
\frac{1}{4}{\cal F}_{\alpha \beta}{\cal F}^{\alpha \beta}$
\cite{plebanski1970lectures}.
Variation of   action (\ref{eq:fT1}) with respect to the
  tetrad yields the field
  equations~\cite{Krssak:2015oua}:
\begin{eqnarray}\label{eq:fT}
	&&
 \zeta_a{}^\mu\equiv\frac{1}{4}f(T) h_a{}^\mu +  f_T \left[ T^b{}_{\nu 
a} S_b{}^{\mu \nu } +
\frac{1}{h}\partial_{\nu}(h S_a{}^{\mu \nu }) \right]   +  f_{TT}\
S_a{}^{\mu\nu} \partial_\nu T- \frac{1}{2}\kappa^2 \Theta_a{}^\mu=0\,,
\end{eqnarray}
with  $f_T\equiv \partial f/\partial T$ and  $f_{TT}\equiv\partial^2 f/\partial
T^2$, and where the elecromagnetic stress-energy tensor
is
 \begin{equation} \label{max1}
\Theta_a{}^\mu\:={\cal F}_{a \alpha} {\cal F}^{\mu \alpha}-\frac{1}{4}
\delta_a{}^\mu {\cal F}_{\alpha \beta} {\cal F}^{\alpha \beta}. \end{equation}
Moreover, variation of   action (\ref{eq:fT1})  with respect to the Maxwell
field gives
\begin{equation} \label{maxf}
\partial_\nu \left( \sqrt{-g} {\cal F}^{\mu \nu} \right)=0.\end{equation}

 We can rewrite equation (\ref{eq:fT})    purely in terms of
spacetime indices by contracting with $g_{\mu\rho}$ and $h^a{}_\sigma$,
resulting to
\begin{align}
\label{eq:fT2}
 H_{\sigma\rho} = \frac{1}{2}\kappa^2 \Theta_{\sigma\rho}\,.
\end{align}
The  symmetric part of (\ref{eq:fT2}) was sourced by the energy-momentum tensor
(\ref{max1}), while their anti-symmetric part is a vacuum constraint for the
considered matter models. The latter is equal to the variation of the action
with respect to the flat spin-connection components \cite{Golovnev:2017dox,
Hohmann:2017duq}, namely
\begin{align}
	 H_{(\sigma\rho)} = \frac{1}{2}\kappa^2 \Theta_{(\sigma\rho)}, \quad  H_{[\sigma\rho]} =0\,.
\end{align}
The explicit forms of these equations can be seen  in Eqs. (26) and (30) of
\cite{Hohmann:2018rwf} by setting the scalar field $\phi$ to zero, however we
do not display them here since we will derive the spherically
symmetric field equations directly from \eqref{eq:fT}.

We proceed by focusing on spherically symmetric solutions.
Employing the spherical coordinates $(t,r,\theta, \phi)$ we write the suitable
spherically symmetric tetrad space as:
\begin{equation}
\!b^a{}_{\mu}\!=\!\left(
\begin{array}{cccc}\!\!
\sqrt{a} & 0 & 0 & 0 \\
0 & \frac{1}{\sqrt{b}} \cos (\phi ) \sin (\theta ) & r \cos (\phi ) \cos (\theta )  & -r \sin (\phi ) \sin (\theta )  \\
0 & \frac{1}{\sqrt{b}} \sin (\phi ) \sin (\theta )  & r \sin (\phi ) \cos (\theta )  & r \cos (\phi ) \sin (\theta ) \\
0 & \frac{1}{\sqrt{b}} \cos (\theta ) & -r \sin (\theta ) & 0 \\
\end{array}
\right)\label{tet},
\end{equation}
where $a\equiv a(r)$ and $b\equiv b(r)$ are two positive  $r$-dependent
functions. The above tetrad  corresponds to the usual metric
\begin{equation}
ds^2=-a(r) \,dt^2+\frac{dr^2}{b(r)}+r^2d\Omega^2,\label{met1}
\end{equation}
with $ d\Omega^2=(d\theta^2+\sin^2\theta d\phi^2)$.
 Using   (\ref{tet})  the torsion scalar becomes
\begin{align}\label{scaletorsion}
  T= \frac{2 \left[1-\sqrt{b(r)}\right] \left[r a'(r)-a(r)
\sqrt{b(r)}+a(r)\right]}{r^2 a(r) b(r)}\,.
\end{align}
Note that $T$ becomes zero in the case   $a=b\rightarrow 1$.

Inserting the above tetrad choice into  the field equations (\ref{eq:fT}) we
acquire
\begin{eqnarray}
\zeta_t{}^t &=&   \frac{1}{4} f+\frac{b^{3/2}(ra'+2a)+rab'-rba'-2ba}{2 r^2 a
b^2}f_T +   \frac{(\sqrt{b}-1) }{r b}T'
f_{TT}  -\frac{\mathbb{Q}'^2}{2ab}=0\,,\label{Eq1}\\
\zeta_r{}^r&=& \frac{\sqrt{b}(ra'+2a)-2(ra'+a)}{2 r^2 ab}f_{T} + \frac{f}{4}-\frac{\mathbb{Q}'^2}{2ab}=0\,,\label{Eq2}\\
\zeta_\theta{}^\theta&=&\zeta_\phi{}^\phi=\frac{2a\sqrt{b}-(ra'+2a) }{4 r a
b}T'f_{TT}+ \frac{f}{4}+\frac{\mathbb{Q}'^2}{2ab}\nonumber\\
&&\ \ \
\ \ \ \
\ \, +\frac{b(r^2a'^2-6raa'-2r^2aa''-4a^2)+a[(2a+ra')(rb'+4b^{3/2})-4ab^2]}{8
r^2 a^2 b^2}f_{T}
=0\,,\label{Eq3}
\end{eqnarray}
 where  primes denote derivatives with respect to $r$.
In the above equations we have introduced the   components of the electric
field  $\mathbb{Q}_{\mu}=[\mathbb{Q}(r),0,0,0]$, where
   ${\cal
F}_{\mu\nu}=\mathbb{Q}_{\mu,\nu}-\mathbb{Q}_{\nu,\mu}$. Hence, the
non-vanishing components of the Maxwell field are
\begin{equation}
\frac{\mathbb{Q}'[a(rb'-4b)+rba']-2rba\mathbb{Q}''}{2ra^2b^2}=0.\label{Eq4}
\end{equation}
Note that      equations  (\ref{Eq1})-(\ref{Eq4}) coincide
with  those of  \cite{Bahamonde_2019} when $\mathbb{Q}=0$.

In the following we solve the above equations to   first-order
expansion  around the Reissner-Nordstr\"om background, which allows us to
extract analytical solutions (since in general the torsion scalar is not a
constant, in which case one has the simple
Reissner-Nordstr\"om solution). Hence, we assume  the perturbative general
vacuum charged solution as
\begin{align}
    a(r)&= 1 - \frac{2M}{r}+\frac{s^2}{r^2}+\epsilon a_1(r)\,,\label{h}\\
   b(r)&= 1 - \frac{2M}{r}+\frac{s^2}{r^2}+\epsilon b_1(r)\,,\label{k}\\
    \mathbb{Q}(r)&=  -\frac{s}{r}+\epsilon \mathbb{Q}_1(r)\label{q}\,.
\end{align}
Finally, concerning the $f(T)$ function we will
consider the  power-law form
\begin{eqnarray}
 f(T)=T+\frac{1}{2}\alpha \epsilon\, T^2\,,\label{tor}
\end{eqnarray}
  with   $\alpha$   the usual parameter of 
the $T^2$ term and  $\epsilon<<1$  the small tracking parameter used to 
quantify 
the expansion in a consistent  way  \cite{DeBenedictis:2016aze,Bahamonde_2019}. 
The above expression in the limit $\epsilon\rightarrow0$ recovers Teleparallel 
Equivalent of General Relativity, and it is known to be a
good approximation for every realistic  $f(T)$ gravity
\cite{Nesseris:2013jea,Nunes:2016qyp,Li:2018ixg}, since the extra term 
quantifies the deviation from General Relativity.

Substituting (\ref{h})-(\ref{tor}) into
(\ref{Eq1})-(\ref{Eq3}),
keeping $\epsilon$ terms up to   first order,  we obtain:
\begin{widetext}
\begin{eqnarray}
\zeta_t{}^t &=&\frac{\epsilon}{r^8\varrho^2}\Big\{
\alpha(\varrho-1)\left[
(10\varrho^2+5\varrho+1)(\varrho-1)^2r^4+2s^2(8\varrho^2+4\varrho+1)(\varrho-1)r
^2+s^4(3\varrho+1)\right]
-r^4s^2a_1-r^7\varrho^6b'_1
\nonumber\\
&&+(\varrho^2 r^2-2r^2+s^2 )r^4\varrho^4b_1+2\varrho^2r^6
s\mathbb{Q}_1'\Big\}=0 \,,
\nonumber\\
\zeta_r{}^r&=&\frac{\epsilon\left\{\alpha(\varrho-1)
[(\varrho-1)^2r^2+s^2(3\varrho-1)][(\varrho-1)r^2+s^2]+r^7\varrho^2a'_1{}
^2+(\varrho^2-1)r^6a_1+r^6\varrho^4b_1+
2\varrho^2r^2s\mathbb{Q}_1'\right\}}{r^8\varrho^2}=0\,,
\nonumber\\
\zeta_\theta{}^\theta&=&
\zeta_\phi{}^\phi=\frac{\epsilon}{4r^{10}\varrho^4}\Big\{2\alpha\varrho\left[
(5\varrho^2+4\varrho+1)(\varrho-1)^4r^6+s^2(\varrho-1)^2
(8\varrho^3-11\varrho^2-6\varrho-3)r^4+s^4r^2(3-10\varrho^3+7\varrho^2)-s^6
\right]\nonumber\\
&&+2\varrho^4r^{10}a''_1-8\varrho^4r^8s\mathbb{Q}_1'
+\varrho^2r^7[(1-3\varrho^2)r^2+s^2]a'_1-\varrho^6r^7[(\varrho^2+1)r^2-s^2]b'_1
+a_1[r^8(1-\varrho^4)-2r^6s^2+r^4s^4]\nonumber\\
&&+\varrho^4 b_1[(\varrho^4-1)r^8+2r^6s^2-r^4s^4]\Big\}=0\,,\label{Eq7}
\end{eqnarray}
\end{widetext}
while  the Maxwell field equation (\ref{Eq4}) becomes
\begin{eqnarray}
 &&
 \!\!\!\!\!\!\!\!\!\!\!\!\!\!\!\!\!\!\!
 2\varrho^4r^5\mathbb{Q}_1''+4v\varrho^4r^4\mathbb{Q}_1'-s\left\{
\varrho^2r^3a'_1-\varrho^6r^3b'_1
-[(\varrho^2-1)r^2+s^2](a_1-\varrho^2b_1)\right\} =0\,,\label{Eq8}
\end{eqnarray}
 where $\varrho=\sqrt{1-\frac{2M}{r}+\frac{s^2}{r^2}}$.
 We solve the above equations separately in the cases where
$\mathbb{Q}_1(r)=0$ and  $\mathbb{Q}_1(r)\neq 0$, namely the cases with or
without a perturbative correction in the charge profile.

\begin{itemize}
 \item \underline{Case I: $\mathbb{Q}_1(r)=0$:}\\

In this case    differential equations  (\ref{Eq7})
and (\ref{Eq8}) admit the solution:
\begin{eqnarray} \label{Eqs0}
  &&
  \!\!\!\!\!\!\!
  a_1(r)=\frac{1}{s^6r^6\varrho\varrho_1}
  \Bigg\{4sr^5\alpha \varrho\Big\{
5\varrho_1{}^{3/2}r^2\varrho^2\tan^{-1}\Theta
+s\left[
rM(5s^2-6M^2)+2M^4
+s^2(M^2-2s^2)\right]\tan^{-1}\Theta_1\Big\}\nonumber\\
    &&
     \ \,\ \ \ \
    +\varrho_1 \Big\{4\alpha
s^2r^5\varrho\ln\varrho\left[(s^2-3M^2)r+M(M^2+s^2)\right]\nonumber\\
    &&
   \ \,\ \ \ \  \ \  \ \,\ \ \ +
  \varrho
\Big\{2r^6\alpha\ln\varrho_1(2M^4+M^2s^2-2s^4)
 -s^2\Big\{8r^5\alpha\ln(r)\Big[
(s^2-3M^2)r+M^3 +s^2M\Big]
\nonumber\\
  &&\ \ \ \ \ \,\ \ \ \ \ \ \ \ \ \ \ \ -s^2\left[s^2c_2r^6-(12\alpha
M+s^2c_1)r^5
    +\alpha(4r^4\varrho_1{}^2-4/3
s^4 r^2+3/5 s^2)\right]\Big\}\Big\}\nonumber\\
&&  \ \,\ \ \ \  \ \  \ \,\ \ \
    -20r^2s^2\varrho\alpha\Big[r^3(4/3s^2-2M^2) +2/15Ms^4
  +r^2M(M^2-1/3s^2)
+rs^2/3(M^2-1/5s^2)\Big]\Big\}\Bigg\}\,,
      \end{eqnarray}
\begin{eqnarray}
 &&
  \!\!\!\!\!\!\!
    b_1(r)=-\frac{4}{s^6r^2\varrho^{7/2}\varrho_1{}^2}\Bigg\{sr^4\alpha
\varrho^{3/2}
\varrho_1^{5/2}\Big\{5\varrho_1{}^{3/2}[s^2(2M+r)-3rM^2]\tan^{-1}\Theta
\nonumber\\
    &&    \ \,\ \ \ \  \ \  \ \,\ \ \ \ \,\ \ \ \  \ \  \ \,\ \ \ \ \,\ \ \ \  \
\  \ \,\ \ \ \ \,\ \ \ \  \ \  \ \
    +2s\Big
[ (2r+5M)s^4-M^2s^2(6M+11r) +10rM^4\Big]\tan^{-1}\Theta_1\Big\}
    \nonumber\\
    && \ \  \ \ \  \
    +\varrho_1^2\Bigg\{2s^2r^4\alpha
\varrho^{3/2}[3M(M+r)s^2-s^4-5rM^3]\ln
\varrho+r^4\varrho^{3/2}(2M^4+2M^2s^2-s^4)(2Mr-s^2)\alpha
\ln\varrho_1{}^2\nonumber\\
    &&
    \ \ \ \ \  \ \ \  \ \  \
    +s^2\bigg\{4r^4\varrho^{3/2}\alpha \ln(r)[5M^3r-3Ms^2(M+r)+s^4]-8/3\alpha
r s^{10}+1/10\alpha s^8(23\varrho^{3/2}+40Mr^2+100r^3)\nonumber\\
    &&
       \ \ \ \ \  \ \ \  \ \  \  \ \  \  \ \  \, \  \
    +rs^6[28\alpha r^4-r^3/2(\varrho^{3/2}c_2+168\alpha M)+6\alpha
r^2M^2-8/3r\alpha\varrho^{3/2}-4M\alpha \varrho^{3/2}]
 \nonumber\\
    && \ \  \ \ \  \
+r^3s^4\Big\{46/3\alpha
r^4-128Mr^3\alpha +r^2[(Mc_2-c_1/2)\varrho^{3/2}+188M^2\alpha]+8\alpha
r(65M^3/3+4\varrho^{3/2})+
16M\alpha \varrho^{3/2}\Big\}
 \nonumber\\
    && \ \  \ \ \  \
-
2r^4Ms^2\alpha\Big[2M^2r^2-11Mr^3+M\varrho^{3/2}
     +3r^4+16M^3r\Big]+6r^6M^3\alpha (r-2M)^2\bigg\}\Bigg\}\Bigg\},
     \label{Eqs0b}
  \end{eqnarray}
  where $\Theta=\frac{Mr-s^2}{s\varrho}$, $\Theta_1=\frac{M-r}{\varrho_1}$ and
$\varrho_1=\sqrt{M^2-s^2}$.  Expressions (\ref{Eqs0}),(\ref{Eqs0b}) are the
solution of the
field equations (\ref{eq:fT}) and (\ref{maxf}) up to ${\mathcal{O}}(\epsilon)$.

 \item \underline{Case II: $\mathbb{Q}_1(r)\neq 0$:}\\

  In this case the solution of (\ref{Eq7}) and (\ref{Eq8}) in the case
$\mathbb{Q}_1(r)\neq 0$ for the metric functions is
 \begin{eqnarray} \label{Eqa}
   &&
   \!\!\!\!\!\!\!\!\!\!\!\!\!\!\!\!\!
   a_1(r)=c_4+
\frac{1}{r^2\alpha^{5/2}}\Bigg\{\int\frac{1}{r^6}\Bigg\{\alpha\Big[\varrho^{1/2}
r^2(s^2r^4\!-\!r^6\!-\!s^6\!+\!r^2s^2)+2r\varrho(3r^4s^2\!-\!2r^6\!-\!s^6)\!+\!r
^2 \varrho^ { 3/2 } (5s^4\!-\!3r^4\!+\!6r^2s^2)
  \nonumber\\
&& \ \ \ \ \ \ \ \ \  \  \ \ \ \ \
\ \ \ \ \ \ \ \ \
+2r\varrho^2(10s^4\!+\!8r^4\!-\!22r^2s^2)
 +\varrho^{5/2}
(18s^4\!+\!15r^4\!-\!23r^2s^2)\!+\!2r\varrho^3(10r^2\!-\!9s^2)\!+\!3r^2\varrho^{
7/2} \Big ] \nonumber\\
&& \ \ \ \ \ \ \ \ \  \  \ \ \ \ \
\ \ \ \ \ \ \ \ \,
-2r^7s\varrho^ { 5/2}\mathbb{Q}''_1\Bigg \}d    r+c_3\alpha^{5/2}\Bigg\}\,,
    \label{Eqbbb}
    \end{eqnarray}
\begin{eqnarray} \label{Eqb}
  &&  b_1(r)= \frac{1}{r^4\varrho^{5/2}}\Big\{\alpha r^2\varrho^{5/2}+4r\alpha
\varrho^2(s^2-r^2)+\varrho^{3/2}[2r^6s\mathbb{Q}'_1+r^7a'_1+r^6a_1+6\alpha
r^4-10r^2s^2\alpha+3\alpha s^4]\nonumber\\
    &&
    \ \ \ \ \ \ \ \ \ \ \ \ \ \ \ \ \ \ \ \ \
    -4r\alpha \varrho(r^2-s^2)^2 -r^2\varrho^{1/2}[2r^2\alpha
s^2+r^6a_1-\alpha s^4-\alpha r^4]\Big\}\,,
    \end{eqnarray}
    while inserting these into  (\ref{Eq8}) we finally acquire
\begin{eqnarray}
&&
\!\!\!\!\!\!\!\!\!
\mathbb{Q}_1(r)=
\frac{1}{15r^{5}\varrho \varrho_1 s^6}\Bigg\{30\alpha r^4s\varrho
\varrho_1 [4M^3r-2Mrs^2-3M^2s^2+s^4]\ln(r^2\varrho^2) \nonumber\\
 &&
 \ \ \ \ \ \ \
 +30\alpha r^4s\varrho
\tanh^{-1}\left(\frac{M-r}{\varrho_1}\right)[(8M^4
-8M^2s^2+s^4)r+Ms^2(5s^2-6M^2)]\nonumber\\
 &&
  \ \ \ \ \ \ \
 +75\alpha r^4\varrho\varrho_1{}^{5/2}
\tanh^{-1}\left(\frac{Mr-s^2}{sr\varrho}
\right)(5M^2r-4Ms^2-s^2r)+300\alpha r^4\varrho_1{}^2\ln2
[(5M^2-\epsilon^2)r-4M\epsilon^2]\nonumber\\
 &&
   \ \ \ \ \ \ \
 -15r^5M\alpha s\varrho
\varrho_1\ln\varrho_1{}^2(6M^2-5s^2) -60r^4\alpha s \varrho
\varrho_1\ln\,r\Big[4M^3r-2Mrs^2 -3M^2s^2+s^4\Big]
 \nonumber\\
 && \ \ \ \ \ \  \
 +s\varrho_1\Bigg\{
\varrho\left[
15(rc_4-c_3)r^4s^5+2s^2\alpha(3s^6+15r^4s^2-10s^4r^2+30Ms^2r^3-60M^2r^4)\right]
\nonumber\\
 &&\ \ \ \ \,\ \ \ \ \ \ \ \ \ \  \
 +5s^4r\alpha(25r^4-125Mr^3-2s^4+23s^2r^2-15r^2M^2-6s^2rM) \nonumber\\
 &&\ \ \ \ \,\ \ \ \   \ \ \ \ \ \  \ +25\alpha
r^5M^3(29r-30M)+25Ms^2r^4\alpha(29M^2-13r^2
-19Mr)\Bigg\}\Bigg\}\,.
   \end{eqnarray}

    \end{itemize}

 Hence, we have extracted the spherically symmetric solutions in the case 
of a quadratic deviation from Teleparallel Equivalent of General Relativity, 
which as we mentioned above is the first correction in every realistic $f(T)$ 
gravity. We stress here that all the above expressions  in the limit 
$\epsilon\rightarrow0$ recover the Schwarzschild results, hence our solutions 
provide the corrections on the latter brought about the  $f(T)$ gravity.

\section{  Properties of the solutions  }\label{S4}

In this section we examine the properties of the extracted perturbative
solutions, and in particular their asymptotic forms, the invariants, and their
energy.

\subsection{Asymptotic forms}

In the case  $\mathbb{Q}_1(r)=0$, the asymptotic form of the solutions
(\ref{Eqs0}),(\ref{Eqs0b}), up to ${\mathcal{O}}(\epsilon)$ become
\begin{eqnarray} \label{mets1}
&&
\!\!\!\!\!\!\!\!\!\!\!
a(r)\approx 1-\frac{2M}{r}+\frac{s^2}{r^2}
\nonumber\\
&& \ \ \ \,
+\epsilon\Bigg\{c_2-\frac{c_1}{r}
-\frac { \alpha}{r}\Bigg[\frac{68}{3M}+\frac{10M^2}{s^2{\varrho_1}}
-\frac{6}{{\varrho_1}}-\frac{76M}{3s^2}+
\frac{20\ln(2M-2s)}{s}-\frac{40M^2\ln(2M-2s)}{s^3}\Bigg]+{\mathcal{O}}\Big(\frac
{1}{r^5}\Big)\Bigg\}\,,
\end{eqnarray}
\begin{eqnarray}
&&
\!\!\!\!\!\!\!\!\!\!\!
b(r)\approx1-\frac{2M}{r}+\frac{s^2}{r^2}
+\epsilon
\Bigg\{\frac{c_1}{r}\!-\!\frac{
2Mc_2}{r}\!+\!\frac{\alpha}{r}\Bigg[\frac{68}{3M}\!+\!\frac{10M^2}{s^2{\varrho_1
}}
\!-\!\frac{6}{{\varrho_1}}\!-\!\frac{76M}{3s^2}+
\frac{20\ln(2M\!-\!2s)}{s}\!-\!\frac{40M^2\ln(2M\!-\!2s)}{s^3}\Bigg]\nonumber\\
&& \ \ \ \,
-\frac{4Mc_1\!+\!(s^2\!-\!8M)s^2}{r^2}\!-\!\frac{8\alpha}{3s^3r^2\varrho_1}
\left[ 30M(2M^2\!-\!s^2)\varrho_1\ln(2\varrho_1{}^2)\!-\!Ms(15M^2\!-\!9s^2)
\!-\!2s\varrho_1(17s^2\!-\!18M^2)\right]\Bigg\}\,,
\label{mets1b}
\end{eqnarray}
and thus the metric (\ref{met1}) becomes Minkowski for $r\rightarrow
\infty$.

On the other hand, in the case $\mathbb{Q}_1(r)\neq0$, the asymptotic forms of
the solutions  (\ref{Eqbbb}),(\ref{Eqb})  up to ${\mathcal{O}}(\epsilon)$
become
\begin{eqnarray} \label{mets2}
&&
\!\!\!\!\!\!\!\!\!\!\!\!\!\!\!\!\!\!\!\!\!\!\!\!\!\!\!\!\!
a(r)=1-\frac{2M}{r}+\frac{s^2}{r^2}
\nonumber\\
&&  \!\!\!\!\!\!\!\!\!
+\epsilon\Bigg\{c_3\Big(1+\frac{2}{s}\Big)+\frac{c_4}{r}-\frac{\alpha}{r}
\Bigg[\frac{8M^3\ln(r)}{s^4}-\frac{8M\ln(r)}{s^2}
-\frac{136M^3}{3s^2\varrho_1{}^2}+\frac{20M^5}{s^4\varrho_1{}^2}+\frac{76M}{
3\varrho_1{}^2}\nonumber\\
&&
\ \ \ \ \ \ \ \ \ \ \ \ \ \ \ \ \ \ \ \ \ \ \ \ \ \ \ \ \ \ \,
+2\tanh^{-1}\Big(\frac{M}{s}\Big) \Bigg(\frac{32M^2}{s^3}-\frac{20M^4}{
s^5}-\frac{12}{s}
\Bigg)\Bigg]+\frac{2sc_3}{r^2}+{\mathcal{O}}\left(\frac{1}{r^3}
\right)\Bigg\}\,,
\end{eqnarray}
\begin{eqnarray}
&&
\!\!\!\!\!\!\!\!\!\!\!\!\!\!\!\!\!\!\!\!\!
b(r)=1-\frac{2M}{r}
+\frac{s^2}{r^2}\nonumber\\
&&  \!\!\!\!
-5\epsilon\Bigg\{\frac{12Mc_3+3c_4s}{rs}-\frac
{\alpha}{r}\Bigg[\frac{60M^3}{s^4}+\frac{24M^3\ln(r)}{s^4}
-\frac{76M}{s^2}-4\tanh^{-1}\Big(\frac{M}{s}\Big)\Bigg(\frac{24M^2}{s^3}
 +\frac{9}{s}+\frac{15M^4}{s^5}
\Bigg)\Bigg]
\nonumber\\
&&   \ \ \, \ \ \ +\frac{\alpha}{3s^5r^2}\Bigg[
240sM^4-208M^2s^3-120M^2s^3\ln(r)+36s^4-48\tanh^{-1}
\Big(\frac{M}{s}\Big)\Big(3Ms^4-10M^3s^2+5M^5
\Big)\Bigg] \nonumber\\
&&   \ \ \, \ \ \
-\frac{(16M^2-s^3)c_3+4c_4sM}{sr^2}+{\mathcal{O}}\left(\frac{1}{r^3}
\right)\Bigg\} \ ,
\label{mets2bb}
\end{eqnarray}
which also become Minkowski in the limit
$r\rightarrow \infty$.

\subsection{Invariants  }

Let us examine the behavior of various invariants in the obtained solutions.
For the case  $\mathbb{Q}_1(r)=0$, and inserting the  asymptotic forms
(\ref{mets1}),(\ref{mets1b}) into the tetrad  (\ref{tet}) and metric
(\ref{met1}), and then
into the various tensor definitions  we respectively acquire the following
expressions for  the torsion tensor square,  the torsion vector square, the
torsion scalar, the Kretschmann scalar,  the Ricci tensor square, and the Ricci
scalar:
\begin{eqnarray}\label{inv1}
&&
 \!\!\!\!\!\!\!\!\!\!\!\!\!\!\!\!\!\!\!
T^{\mu \nu \lambda}T_{\mu \nu \lambda} 
=\frac{16M+8\epsilon(c_1-2Mc_2)}{r^3}-\frac{16s^2-4M^2+\epsilon[68M^2\epsilon 
c_2{}^2-16s^2c_2-4Mc_1+15c_1{}^2\epsilon+8M^2c_2-64Mc_1c_2\epsilon]}{2r^4}
  \nonumber\\
&& \ \ \ \ \ \ \
 +{\mathcal{O}}\Big(\frac{1}{r^5}\Big)\,,
  \end{eqnarray}
\begin{equation}
 \!\!\!  \!\!\!  \!\!\!  \!\!\!  \!\!\!  \!\!\!  \!\!\!  \!\!\!  \!\!\!
 \!\!\!  \!\!\!  \!\!\!  \!\!\!  \!\!\!  \!\!\!  \!\!\!  \!\!\!  \!\!\!  \!\!\!
 \!\!\!  \!\!\!  \!\!\!   \!\!\! \!\!\! \!\!\! \!\!\! \!\!\! \!\!\! \!\!\!
\!\!\! \!\!\! \!\!\! \!\!\! \!\!\! \!\!\! \!\!\!  \!\!\!  \!\!\!  \!\!\!  \!\!\!
 \!\!\!  \!\!\!  \!\!\!  \!\!\!  \!\!\!  \!\!\!
 \!\!\!  \!\!\!  \!\!\!    \!\!\!  \!\!\!
 \!\!\!  \!\!\!  \!\!\!
\!\!\! \!\!\! \!\!\! \!\!\! \!\!\! \!\!\! \!\!\!
 T^\mu T_\mu =
  \frac {4(2M+\epsilon [c_1-2
Mc_2+\epsilon c_1c_2-2M\epsilon 
c_2{}^2])}{r^3}+{\mathcal{O}}\Big(\frac{1}{r^4}\Big)\,,
  \end{equation}
\begin{eqnarray}
&& \!\!\!\!\!\!\!\!\!\!\!\!\!\!\!\!\!\!\!
T(r)=\frac{4c_2\epsilon^2(2Mc_2-c_1)}{r^3} + 
\frac{4M^2-8s^2-\epsilon(8M^2c_2-8s^2c_2-4Mc_1)+\epsilon^2[9c_1{}^2+76M^2c_2{}
^2-16s^2c_2{}^2-56Mc_1c_2]}{2r^4}
  \nonumber\\
&& \ \ \ \ \ \ \
+{\mathcal{O}}\Big(\frac{1}{r^5}\Big)\,,
  \end{eqnarray}
  \begin{eqnarray}
&&\!\!\!  
  \!  \!\!\!  \!\!\!  \!\!\!  \!\!\!  \!\!\!  \!\!\!  \!\!\!  \!\!\!
\!\!\!   \!\!\!   \!\!\!
 \!\!\!  \!\!\!  \!\!\!   \!\!\!
 \!\!\!  \!\!\!  \!\!\!    \!\!\!  \!
 \!\!\!  \!\!\!  \!\!\!  R^{\mu \nu \lambda \rho}R_{\mu \nu \lambda \rho}=
\frac{48M^2+48\epsilon\,M(c_1-2c_2M)+12\epsilon^2(c_1{}^2-6Mc_1c_2+16M^2c_2{}^2)
}{r^6}+{\mathcal{O}}\Big(\frac{1}{r^7}\Big)\,,
\end{eqnarray}
\begin{equation}\!\!\!  
 R^{\mu \nu}R_{\mu 
\nu}=\frac{4\epsilon^2c_2s^2[c_1-2Mc_2]}{r^7}+\frac{2s^2[2s^2-4s^2\epsilon\,
c_2-\epsilon^2(2c_1{}^2-13Mc_1c_2-8s^2c_2{}^2+18M^2c_2{}^2)}
{r^8}+{\mathcal{O}}\Big(\frac{1}{r^9}\Big)\,,  \end{equation}
\begin{eqnarray} &&\!\!\!\!\!   
R=-\frac{\epsilon^2(10M^2c_2{}^2-2s^2c_2{}^2-9Mc_1c_2+2c_1{}^2)}{r^4}-\frac{
\epsilon^2(48M^3c_2{}^2-24s^2Mc_2{}^2-44M^2c_1c_2+10Mc_1{}^2+
  11s^2c_1c_2)}{r^5}
  \nonumber\\
&& 
\frac{-\epsilon^2(168M^4c_2{}^2-128s^2Mc_2{}^2+10s^2c_2{}
^2-156M^3c_1c_2+81Ms^2c_1c_2+36M^2c_1{}^2-10s^2c_1{}^2)}{r^6} \nonumber\\
&& 
 -\frac{\epsilon^2}{r^7}\biggl(112M^3c_1{}^2+512M^5c_2{}
^2+372M^2s^2c_1c_2-16M^3\alpha c_2+60Ms^2c_2{}^2+96Ms^4c_2{}^2  \nonumber\\
&&   
-31s^4c_1c_2-520M^3q^2c_2{}^2-480M^4c_1c_2\biggr)+\frac{16\epsilon\alpha 
M^3}{r^7}+{\mathcal{O}}\Big(\frac{1}{r^8}\Big)\,.
  \end{eqnarray}  
   Similarly, for the case  $\mathbb{Q}_1(r)\neq0$  we obtain the same
expressions, and the only difference is in the  torsion tensor square, which
now becomes
\begin{eqnarray}\label{inv1}
&&   \!\!\!    \!\!\!    \!\!\!    \!\!\!    \!\!\!
 \!\!\!    \!\!\!    \!\!\!    \!\!\!    \!\!\!   \!\!\!   T^{\mu \nu
\lambda}T_{\mu \nu \lambda} =
 \frac{32\epsilon \alpha (3s^4-8M^2s^2+5M^4)\tanh^{-1}
\Big(\frac{M}{s}\Big) }{ r^3s^5}\nonumber\\
&&
+ \frac{ 24s^4[\epsilon(2Mc_1+c_2)-2M]+32\epsilon
M(3c_1s^3+19\alpha s^2-15\alpha M^2)}{3r^3s^4}
 +{\mathcal{O}}\Big(\frac{1}{r^4}\Big)\,.
  \end{eqnarray}
  The above invariants reveal   the presence of the singularity   at $r=0$ as
expected, which is more mild than the case of simple TEGR, a known feature of
higher-order torsional theories \cite{Cai:2015emx, Nashed:2020kjh,
Bahamonde_2019,Ren:2021uqb}.

\subsection{Energy}

One of the advantages of teleparallel formulation of gravity is the easy
handling of the energy calculations, which is not the case  in usual curvature
formulation \cite{Cai:2015emx}. We start with the gravitational
energy-momentum, $P^a$,  which in  integral form in
four dimensions is \cite{ULHOA_2013}
\begin{eqnarray} \label{en}
P^a=-\int_V d^3x \partial_i\Pi^{ai}f_T,\end{eqnarray}
where $V$ is the three-dimensional volume  and $\Pi^{ai}=-4\pi
S^{a 0 i}$ is expressed in terms of the superpotential components.  In the  TEGR
limit,  namely for  $f_T=1$, the above expression reduces to the form given in
\cite{PhysRevD.65.124001}.

We start with the case $\mathbb{Q}_1(r)=0$. Inserting the tetrad functions
(\ref{mets1}),(\ref{mets1b}) into the tetrad  (\ref{tet}) we can calculate the
involved
superpotential component as
 \begin{equation}\label{sup}
S^{001}=\frac{6M^2s^3(\epsilon c_2-1)+\epsilon sM(76M\alpha-3c_1s^2-68\alpha
s^2)+60M\epsilon \alpha
\ln(2(M-s)(2M^2-s^2)}{6Mr^2s^3(1+c_2\epsilon)}+\frac{6\epsilon\alpha(3s^2-5M^2)}
{s^2(1+\epsilon c_2)\varrho_1}\,,
\end{equation}
which substituted into   (\ref{en}) leads to
\begin{eqnarray} \label{en11}
&&
\!\!\!\!\!\!\!\!\!\!\!\!\!\!\!\!\!\!\!\!\!\!\!\!\!\!\!\!\!\!\!\!\!\!\!\!\!\!\!\!
\!\!\!\!\!\!\!\!\!\!\!\!\!\!\!\!\!\!\!\!
{P^0=E\approx
M+\epsilon\alpha\left\{M\frac{15M-38\varrho_1}{3s^2\varrho_1}-20\frac{
(2M^2-s^2)\ln[2(M-s)]}{s^3}\right\}-\frac{M^2+s^2}{2r}}\nonumber\\
&&
\!\!\!\!\!\!
{- 5\epsilon\alpha\left\{M\frac{15M-38\varrho_1}{3s^2\varrho_1 r
}+20\frac{(2M^2-s^2)\ln[2(M-s)]}{r s^3}\right\}={\cal
M}+O\left(\frac{1}{r}\right)}\,,
\end{eqnarray}
where ${{\cal M}\approx M-\epsilon\alpha \frac{38M}{3s^2}}$ is the
Arnowitt-Deser-Misner (ADM) mass
that contains $M$ and $ \epsilon$  up to first order.

In the case $\mathbb{Q}_1(r)\neq 0$, the above procedure leads to
\begin{equation}\label{sup1}
S^{001}=-\frac{M}{r^2}+\epsilon
\left[\frac{12\alpha(5M^4-8M^2s^2+3s^4)
\tanh^{-1}\left(\frac{M}{s}\right)+12Mc_3s^4(s+2)+4sM\alpha(17s^2-15M^2)
+3c_4s^5}{s^5r^2}\right] ,
\end{equation}
and then to
{\small{
\begin{eqnarray} \label{en1}
&&
\!\!\!\!\!\!\!\!\!\!\!\!\!\!\!\!\!\!\!\!\!\!\!\!
P^0=E\approx M+\epsilon\left[\frac{12\alpha
\tanh^{-1}\left(\frac{M}{s}\right)(5M^4-8M^2s^2+3s^4)+
3Mc_3s^4(s+2)
+4sM\alpha(17s^2-15M^2)+3c_4s^5}{3s^5}\right]-\frac{M^2+s^2}{2r}
\nonumber\\
&&\!\!\!\!\!\!\!\!\! \!\!\!\!\!\!\!
-\epsilon\left[\frac{12M\alpha\tanh^{-1}\left(\frac{M}{s}\right)\!
(10M^4\!-\!16M^2s^2\!+\!6s^4)+
3c_3s^4(3M^2s\!+\!6M^2\!-\!2s^2\!+\!s^3)
+8sM^2\alpha(19s^2\!-\!15M^2)
+6c_4Ms^5}{6rs^5}\right],
\end{eqnarray}}}
\begin{eqnarray}
&&
\!\!\!\!\!\!\!\!\! \!\!\!\!\!\!\!\!\!\!\!\!\!\!\!\!
\!\!\!\!\!\!\!\!\!\!\!\!\!\!\!\! \!\!\!\!\!\!\!\!\!\!\!\!\!\!\!\!
\!\!\!\!\!\!\!\!\!\!\!\!\!\!\!\! \!\!\!\!\!\!\!\!\!\!\!\!\!\!\!\!
\!\!\!\!\!\!\!\!\!\!\!\!\!\!\!\! \!\!\!\!\!\!\!\!\!\!\!\!\!\!\!\!
\!\!\!\!\!\!\!\!\!\!\!\!\!\!\!\! \!\!\!\!\!\!\!\!\!\!\!\!\!\!\!\!
\!\!\!\!\!\!\!\!\!\!\!\!\!\!\!\! \!\!\!\!\!\!\!\!\!\!\!\!\!\!\!\!
\!\!\!\!\!\!\!\!\!\!\!\!\!\!\!\! \!\!\!\!\!\!\!\!\!\!\!\!\!\!\!\!
\!\!\!\!\!\!\!\!\!\!\!\!\!\!\!\! \!\!\!\!\!\!\!\!\!\!\!\!\!\!\!\!
\!\!\!\!\!\!\!\!\!\!\!\!\!\!
{ E\approx {\cal M}_1+O\left(\frac{1}{r}\right)}\,,
\end{eqnarray}
 where  ${ {\cal
M}_1=M+\epsilon\left[\frac{68\alpha\,M+3Mc_3s(s+2)}{3s^2}\right]}$. Note that
for  $\epsilon=0$, i.e. in
the TEGR limit, we recover the well-know energy expression of
Reissner-Nordstr\"om  spacetime \cite{NASHED_2007}.

\section{ Geodesic deviation and stability of motion}\label{S6336b}

In this section we proceed to the examination of the  stability of 
motion around   the obtained
black hole solutions, investigating the geodesic deviation.
 The geodesic equations of a test particle in the gravitational field are given by
 \begin{equation}\label{ge}
 {d^2 x^\alpha \over d\lambda^2}+ \Bigl\{^\alpha_{ \beta \rho} \Bigr \}
 {d x^\beta \over d\lambda}{d x^\rho \over d\lambda}=0,
 \end{equation}
where $s$ denotes  the affine  connection parameter and $\Bigl\{^\alpha_{ \beta
\rho} \Bigr \}$ the Levi-Civita connection. The
  geodesic  deviation equations  acquire the form
\cite{1992ier..book.....D}
  \begin{equation} \label{ged}
 {d^2 \psi^\sigma \over d\lambda^2}+ 2\Bigl\{^\sigma_{ \mu \nu} \Bigr \}
 {d x^\mu \over d\lambda}{d \psi^\nu \over d\lambda}+
\Bigl\{^\sigma_{ \mu \nu} \Bigr \}_{,\ \rho}
 {d x^\mu \over d\lambda}{d x^\nu \over d\lambda}\psi^\rho=0\,,
 \end{equation}
where  $\psi^\rho$ is the 4-vector deviation.

In the case of the spherically symmetric ansatz  (\ref{met1}) the above
expressions give
 \begin{eqnarray}
 &&
{d^2 t \over d\lambda^2}=0,\nonumber\\
&& {1 \over 2} a'(r)\left({d t \over
d\lambda}\right)^2-r\left({d \phi \over d\lambda}\right)^2=0, \nonumber\\
&&{d^2
\theta \over d\lambda^2}=0,\nonumber\\
&& {d^2 \phi \over d\lambda^2}=0,
\end{eqnarray}
and therefore for the geodesic deviation  we finally  obtain
\begin{eqnarray}\label{ged11}
&&  {d^2 \psi^1 \over d\lambda^2}+b(r)a'(r) {dt
\over d\lambda}{d
\psi^0 \over d\lambda}-2r b(r) {d \phi \over d\lambda}{d \psi^3 \over
d\lambda}+\left\{{1 \over 2}\left[a'(r)b'(r)+b(r) a''(r)
\right]\left({dt \over d\lambda}\right)^2-\left[b(r)+rb'(r)
\right] \left({d\phi \over d\lambda}\right)^2 \right\}\psi^1=0\,, \nonumber\\
&&  {d^2 \psi^0 \over
d\lambda^2}+{b'(r) \over b(r)}{dt \over d\lambda}{d \psi^1 \over
d\lambda}=0\,,\nonumber\\
&&  {d^2 \psi^2 \over d\lambda^2}+\left({d\phi \over d\lambda}\right)^2
\psi^2=0\,, \nonumber\\
&&   {d^2 \psi^3 \over d\lambda^2}+{2 \over r}{d\phi \over d\lambda} {d
\psi^1 \over d\tau}=0\,.
\end{eqnarray}
Using
the circular orbit
$\theta={\pi \over 2}$, $
{d\theta \over d\lambda}=0$, and ${d r \over d\lambda}=0,
$
we acquire
$
\left({d\phi \over d\lambda}\right)^2={a'(r)
\over r[2a(r)-ra'(r)]}$ and
$
\left({dt \over
d\lambda}\right)^2={2 \over 2a(r)-ra'(r)}$, and thus equations (\ref{ged11})
can be rewritten as
\begin{eqnarray}
\label{ged22} &&  
 {d^2 \psi^1 \over d\phi^2}+a(r)a'(r) {dt \over
d\phi}{d \psi^0 \over d\phi}-2r a(r) {d \psi^3 \over
d\phi} +
\left\{{1 \over 2}\left[a'^2(r)+a(r) a''(r)
\right]\left({dt \over d\phi}\right)^2-\left[a(r)+ra'(r)
\right]  \right\}\psi^1=0\,, \nonumber\\
&& {d^2 \psi^2 \over d\phi^2}+\psi^2=0\,,  \nonumber\\
&&  {d^2 \psi^0 \over d\phi^2}+{a'(r) \over
a(r)}{dt \over d\phi}{d \psi^1 \over d\phi}=0\,,  \nonumber\\
&&  {d^2 \psi^3 \over
d\phi^2}+{2 \over r} {d \psi^1 \over
d\phi}=0\,. \end{eqnarray}
The second equation  of (\ref{ged22}) corresponds to a simple harmonic motion,
which  indicates that  the plane  $\theta=\pi/2$ is stable.
Moreover, the other equations of (\ref{ged22}) have   solutions of the
  form:
\begin{equation} \label{ged33}
  \psi^0 = \zeta_1 e^{i \sigma \phi}\,, \qquad  \psi^1= \zeta_2e^{i 
\sigma
\phi}\,, \qquad {\text{and}} \qquad  \psi^3 = \zeta_3 e^{i \sigma 
\phi}\,,
\end{equation}
where $\zeta_1, \zeta_2$ and $\zeta_3$ are constants.  Substituting
(\ref{ged33}) into
(\ref{ged22}),  we extract
the stability of motion condition   as:
\begin{equation} \label{con1}
 \frac{3abb'-\sigma^2a b'-2rb^{3/2}a'^{3/2}-rab'^2+rab'a'+rab a''}{a b'}>0.
\end{equation}
 Equation (\ref{con1})   has the following solution in terms of the metric
potentials
\begin{equation} \label{stab1}
\sigma^2= \frac{3a bb'-2rb^{3/2}a'^{3/2}-rab'^2+rab'a'+rab a''}{a^2
b'^2}>0.\end{equation}
Hence, in order to conclude on the  stability of motion around   the 
obtained black-hole
solutions,   for the case  $\mathbb{Q}_1(r)=0$  in the above expressions we
insert
  $a(r)$ and $b(r)$  from    (\ref{mets1}),(\ref{mets1b}), while for the case
$\mathbb{Q}_1(r)\neq 0$   from  (\ref{mets2bb}).
\begin{figure}[ht]
\centering
  \includegraphics[scale=0.3]{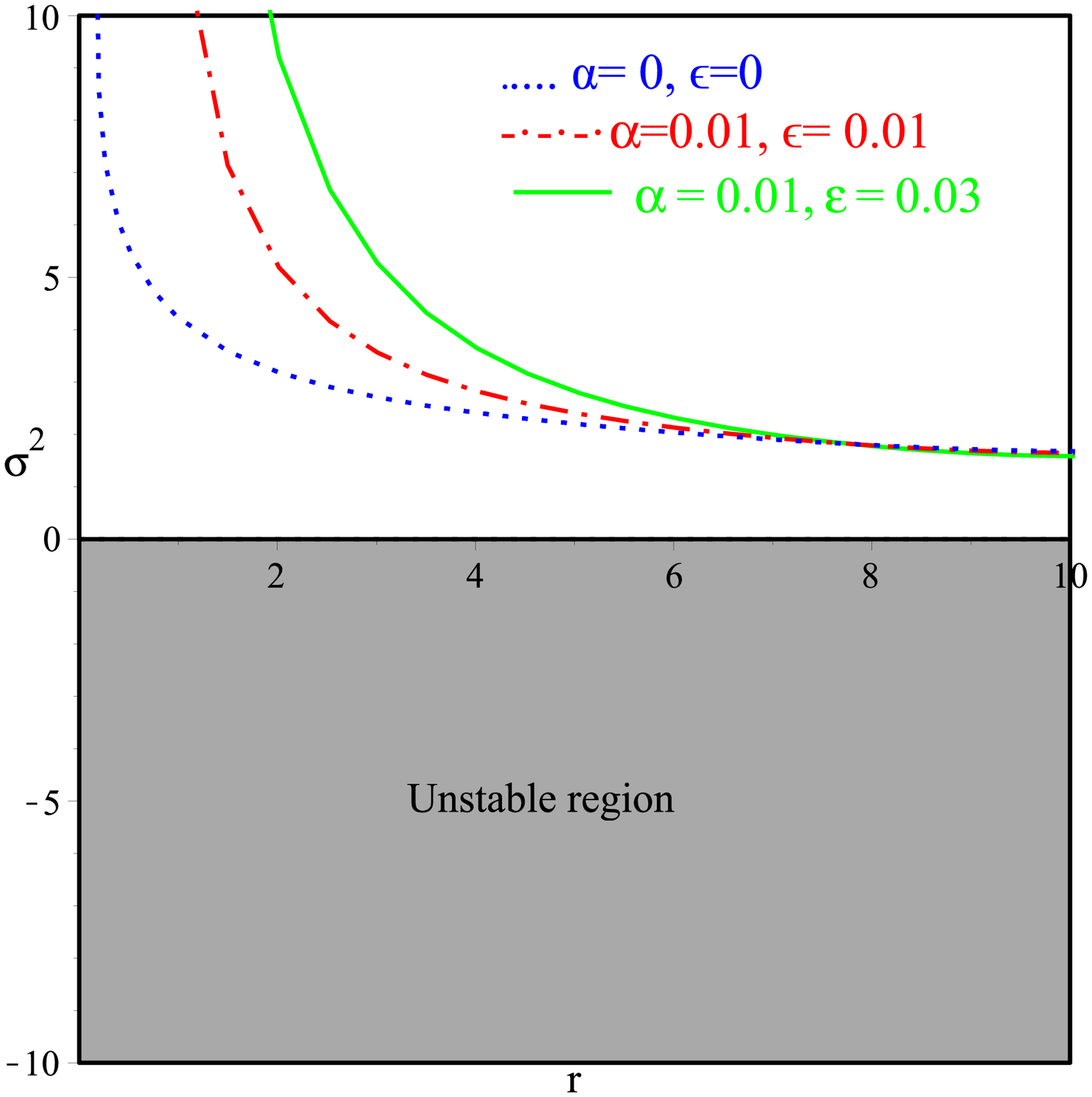}
 \includegraphics[scale=0.3]{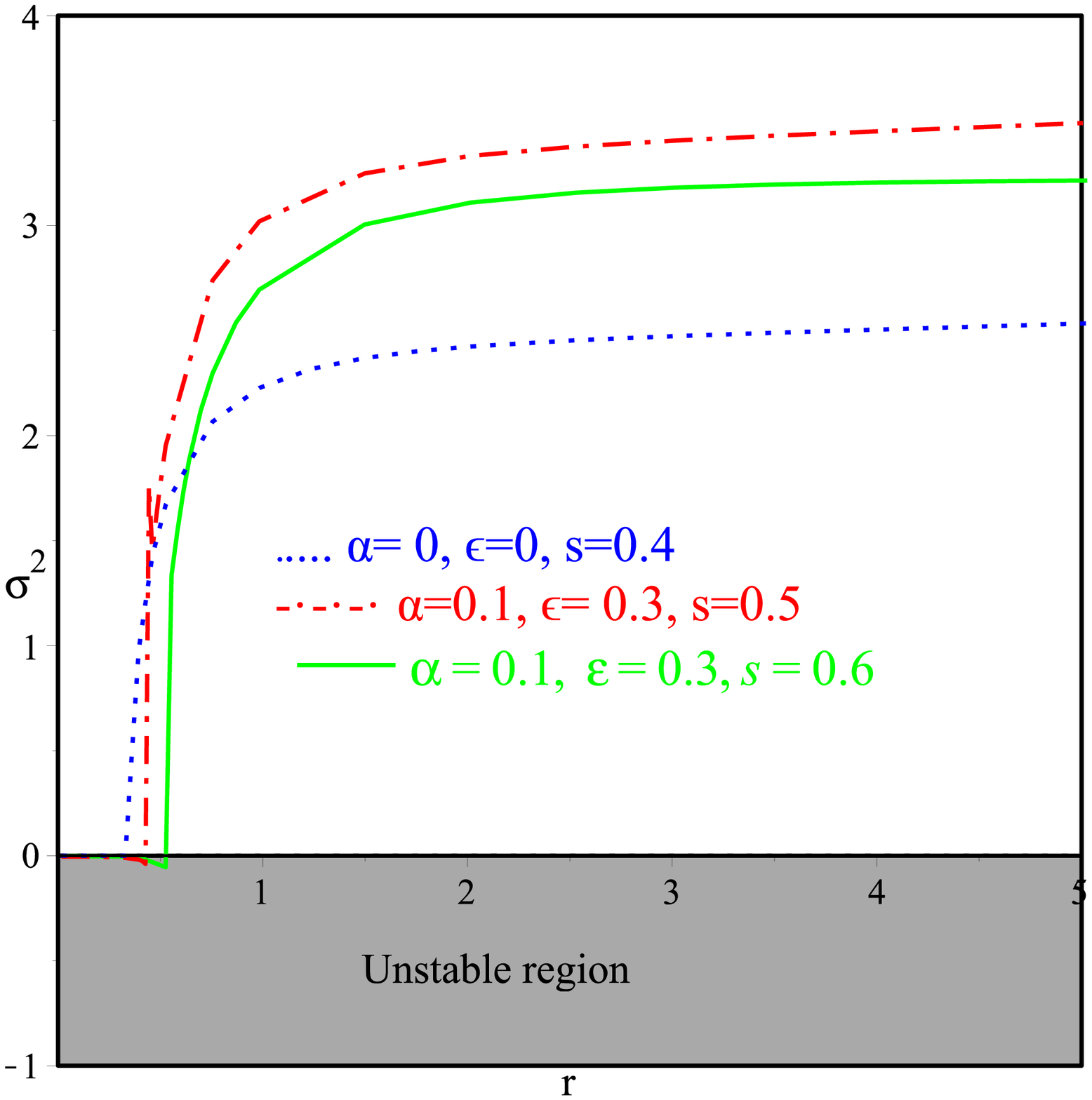}
\caption{ {\it{  The black-hole stability of motion  parameter $\sigma^2$  
versus
$r$. Left graph:  $\mathbb{Q}_1(r)=0$ with  $M=9$, $c_1=c_2=s=1$ and
various choices of the model parameters $\alpha$ and $\epsilon$ in Planck mass
units. Right graph:  $\mathbb{Q}_1(r)\neq 0$ with  $M=0.01$,  $c_1=c_2=1$ and
 various choices of the model parameters $\alpha$, $\epsilon$ and $s$ in Planck
mass
units.}}}
\label{Fig:1}
\end{figure}

In order to present the above results in a more transparent way, in   Fig.
\ref{Fig:1}  we depict the behavior of  $\sigma^2$ for various choices of the
model parameters, for the two cases $\mathbb{Q}_1(r)=0$ and
$\mathbb{Q}_1(r)\neq 0$  separately. Note that for $\mathbb{Q}_1(r)=0$ we
always obtain  stability of motion  as expected, while for $\mathbb{Q}_1(r)\neq 
0$ we find
potentially unstable regions.

\section{Thermodynamics}\label{S5}

In this section we perform an analysis of the thermodynamic properties of the
obtained black-hole solutions. Since the nature of the solutions and especially
their thermodynamic features change for   $\mathbb{Q}_1(r)= 0$ and
$\mathbb{Q}_1(r)\neq 0$, in the following we examine the two cases separately.

\subsection{Thermodynamics of the black hole solution with $\mathbb{Q}_1(r)=
0$}\label{S5a}

\begin{figure}[ht]
\centering
 \includegraphics[scale=0.3]{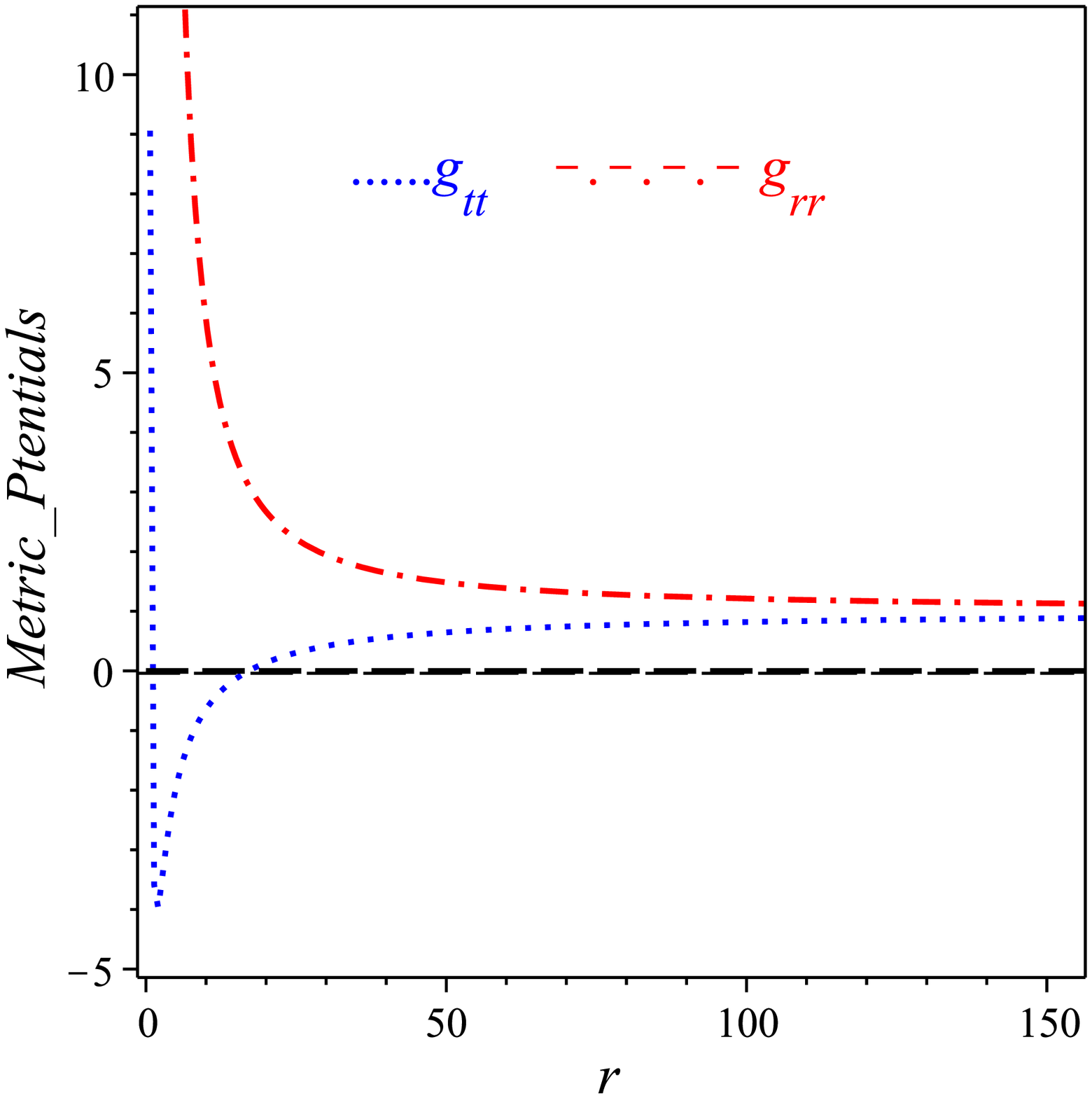}
 \includegraphics[scale=0.3]{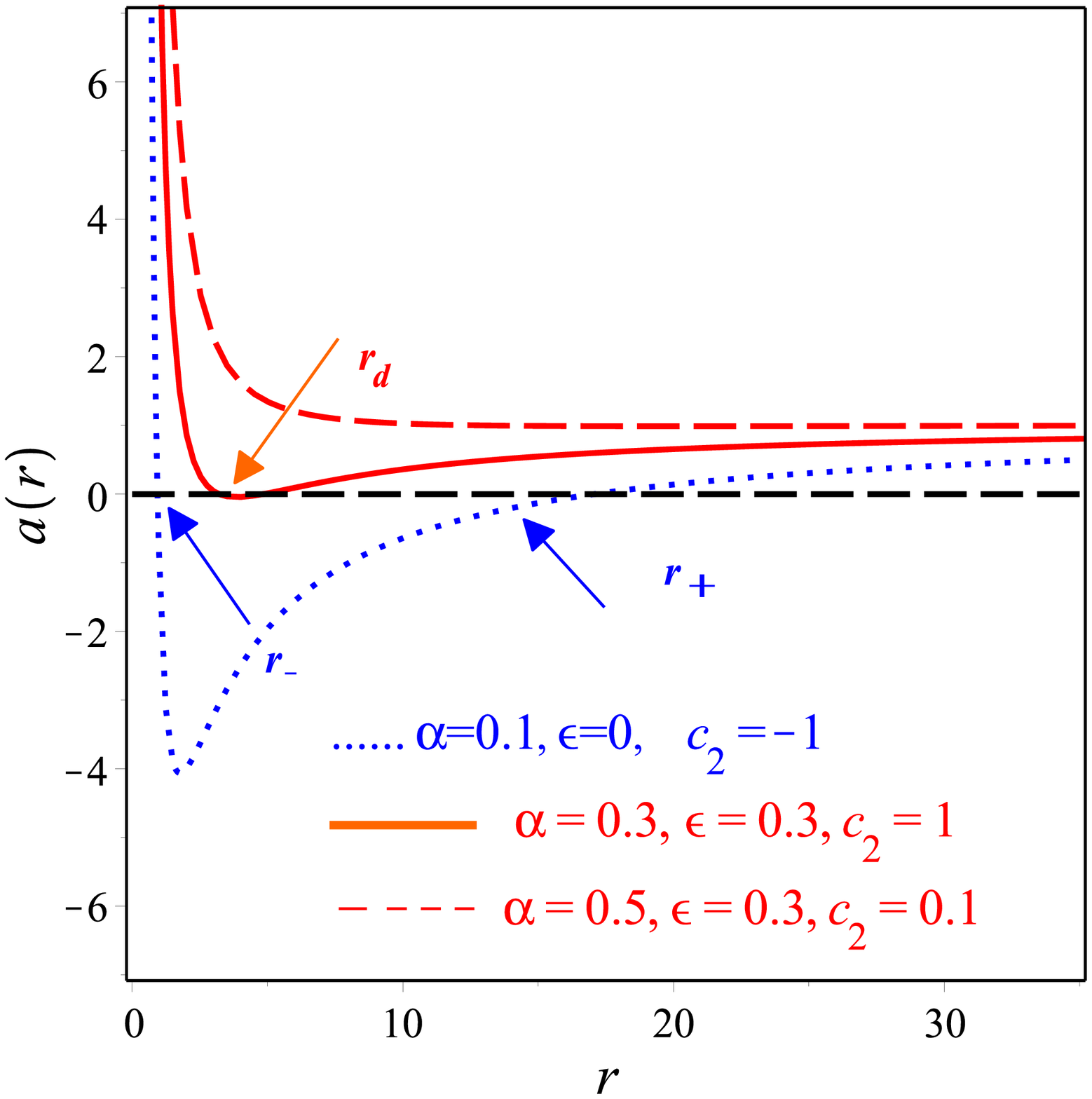}
\caption{{\it{Left graph: The two metric potentials $g_{tt}\equiv a(r)$ and
$g_{rr}\equiv b(r)$ given in
(\ref{mets1}),(\ref{mets1b}) versus $r$, for the black hole solution with
$\mathbb{Q}_1(r)=
0$, for $M=9$, $c_1=c_2=1$, $\alpha=0.1$,
$\epsilon=0.1$,  and
$s=4$, in Planck
mass units. Right graph: The   metric potential  $a(r)$ versus $r$,
for $M=9$, $c_1=1$, $s=4$ and various values of  the model parameters $c_2$,
$\alpha$ and $\epsilon$,   in Planck
mass units.  $r_-$ and  $r_+$ are the inner and outer horizons
respectively, while $r_d $ is the degenerate horizon in which the above two
coincide. }}}
\label{Fig:2}
\end{figure}

We start by investigating the black-hole solution of the case $\mathbb{Q}_1(r)=
0$ given in    (\ref{mets1}),(\ref{mets1b}).
In the left graph of Fig. \ref{Fig:2}  we display the metric
potentials $a(r)$ and $b(r)$. As we can see, $a(r)$ may exhibit two horizons
while $b(r)$ does not. In  the right graph of Fig. \ref{Fig:2} we focus on
$a(r)$, in order to make more transparent the behavior of its possible two
horizons, acquired by solving $a(r) = 0$, namely
  $r_-$ which denotes the  inner Cauchy horizon of the black
hole and $r_+$ which is the outer event horizon. In particular, for small
$\alpha$ values, namely small deviations from general relativity,
 we   obtain  two horizons, however as $\alpha$ increases there is a specific
value in which the two horizons become degenerate ($r_{-}=r_{+}=r_d $), while
for larger values  the horizon disappears and the central singularity
becomes a naked one. This is a known feature of torsional gravity,
namely for some regions of the parameter space naked singularities appear
\cite{Gonzalez:2011dr,Capozziello:2012zj,Nashed:2018cth}.
Finally, let us calculate the total mass contained within the event horizon
$r_+$. We find the  mass-radius expression as
\begin{eqnarray} \label{hor-mass-rad1a}
&&  {{{\cal{M}_+}}\equiv{{\cal{M}(r_+)}} \approx \frac{r_+}{2}\,,}
\end{eqnarray}
 where ${{\cal{M}_+}}$ is given by Eq. (\ref{en11}) for $r_+$ in place of $r$.

\begin{figure}[ht]
\centering
 \includegraphics[scale=0.3]{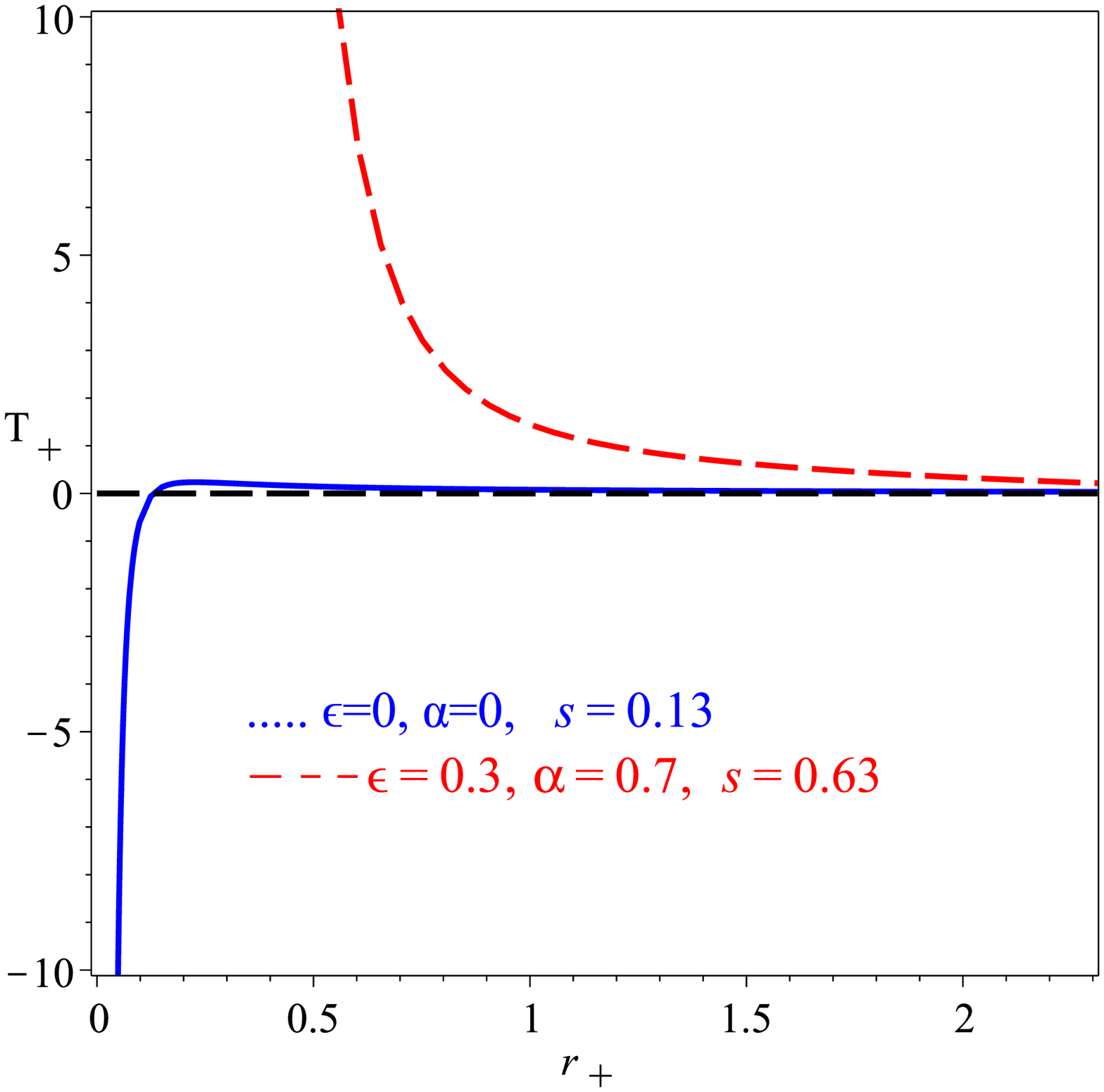}
 \includegraphics[scale=0.3]{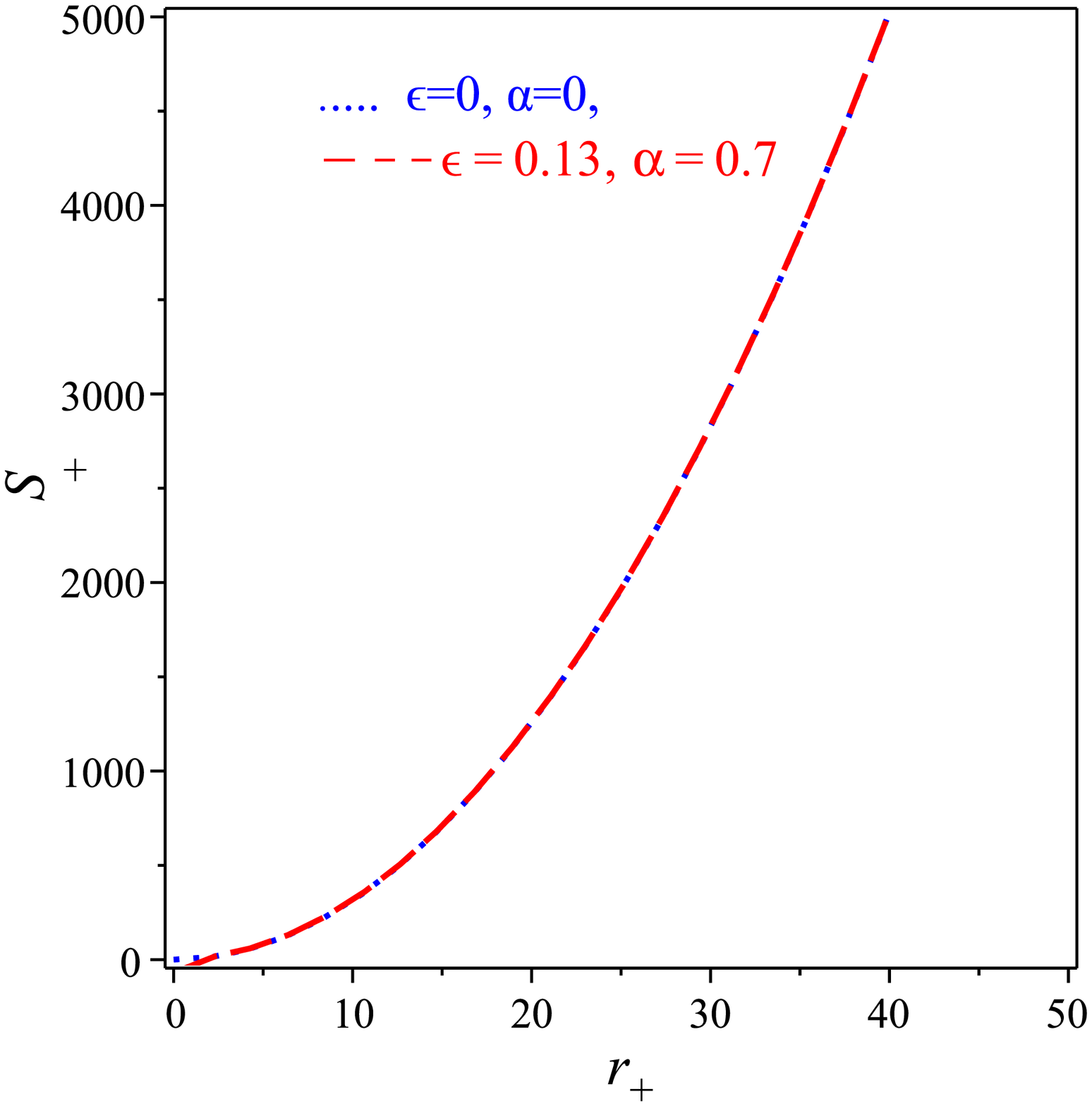}
\caption{{\it{ The temperature  (left graph) and entropy   (right
graph) versus the horizon,
 for the black hole solution with
$\mathbb{Q}_1(r)=
0$, for  $c_1=1$, $c_2=5$ and $s=4$ and for   various values of the model
parameters $\alpha$ and $\epsilon$, in Planck
mass units.
 }}}
\label{Fig:3}
\end{figure}

We proceed by examining the temperature.
The Hawking black-hole temperature is   defined as
\cite{PhysRevD.86.024013,Sheykhi:2010zz,Hendi:2010gq,PhysRevD.81.084040}
  \begin{equation}
T_+ \equiv T(r_+)= \frac{a'(r_+)}{4\pi},
\label{tempdef}
\end{equation}
with $r = r_+$    the event horizon, which satisfies
$a'(r_+)\neq 0$. Additionally, in the framework of   $f(T)$ gravity, the
black-hole entropy   is given by
\cite{Miao:2011ki,PhysRevD.84.023515,Zheng:2018fyn}
\begin{equation}\label{ent}
S_+ \equiv S(r_+)=\frac{A}{4 f_{T}(r_+)},
\end{equation}
where $A$ is the  area. Inserting the $f(T)$ form (\ref{tor})  and the
solution  (\ref{mets1}),(\ref{mets1b}) into the above definitions we find
\begin{equation}\label{m44-ee}
{T_+} \simeq\frac{3r_+{}^2s^3+\epsilon\alpha[30s^2r_++11s^3+30r_+{}
^3\ln(2/r_+{})+45sr_+{}^2]+3\epsilon s^3r_+{}^2c_2}{12\pi r_+{}^3}\,,
\end{equation}
and
\begin{eqnarray} \label{ent1}
{S_+}\simeq \pi
r_+{}^2\left[1+2\epsilon \alpha
\left(\frac{4}{r_+{}^2}+ \frac{\cal{M}_+{}^2-2s^2}{r_+{}^4}\right)\right]\,.
\end{eqnarray}
These expressions
  indicate  that for $\epsilon=0$ we   recover the standard
general-relativity temperature and  entropy.
In   Fig. \ref{Fig:3} we depict the   temperature and  entropy versus the
horizon, for various values of the model parameters. As we can see the entropy
is always positive and exhibits a quadratic behavior, while the temperature is
always positive when $\epsilon>0$ but for    vanishing
$\epsilon$ it is  positive    only for $r_d>r_+$.

\begin{figure}[ht]
\centering
 \includegraphics[scale=0.33]{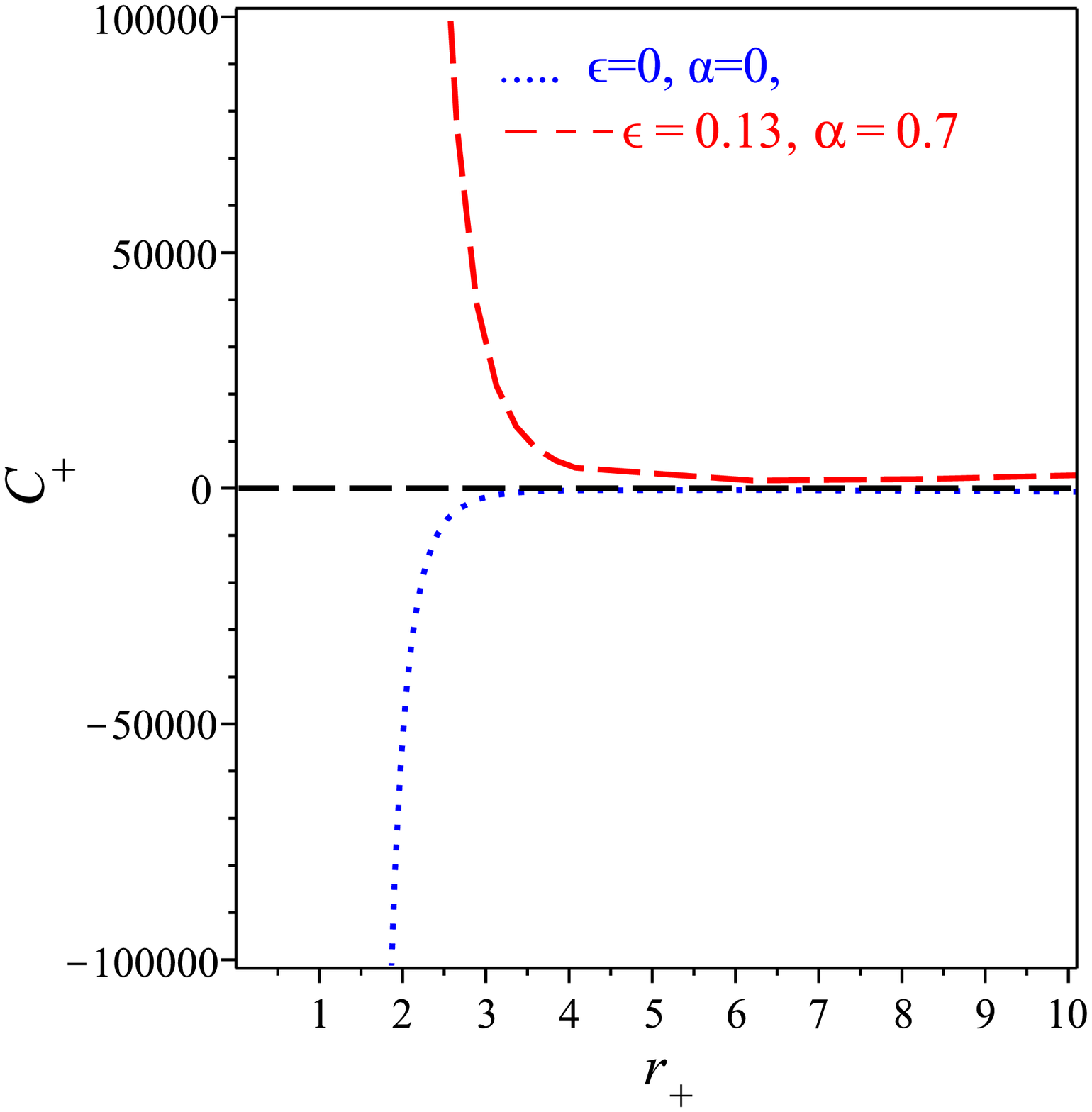}
 \includegraphics[scale=0.33]{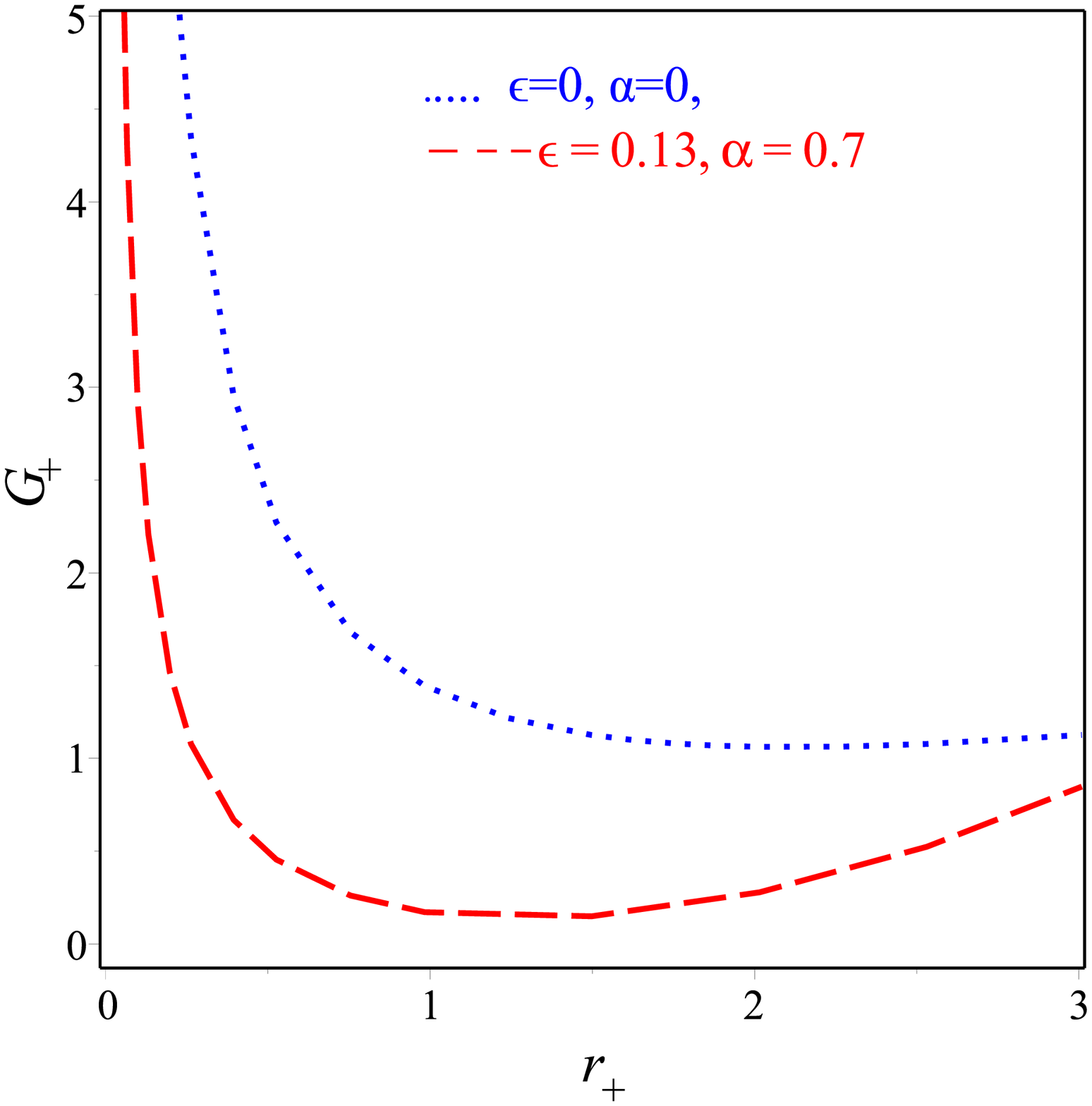}
\caption{{\it{
 The heat capacity  (left graph) and the Gibb's energy  (right
graph) versus the horizon,
 for the black hole solution with
$\mathbb{Q}_1(r)=
0$, for  $c_1=1$, $c_2=5$ and $s=4$ and for   various values of the model
parameters $\alpha$ and $\epsilon$, in Planck
mass units.}}}
\label{Fig:3b}
\end{figure}

We now focus on the  heat capacity, which is a crucial quantity concerning the
thermodynamic stability \cite{Nashed:2003ee,Myung_2011,Myung:2013oca}, since our
perturbative approach to the black-hole solution allows for an easy calculation.
 The  heat capacity at the event horizon
is defined as   \cite{Nouicer:2007pu,DK11,Chamblin:1999tk}:
\begin{equation}\label{heat-capacity}
C_+\equiv C(r_+)\simeq\frac{\partial \cal{M}_+}{\partial T_+}=\frac{\partial
\cal{M}_+}{\partial r_+} \left(\frac{\partial T_+}{\partial r_+}\right)^{-1}\, ,
\end{equation}
and positive  heat capacity implies thermodynamic stability.
Substituting  (\ref{hor-mass-rad1a}) and (\ref{m44-ee}) into
(\ref{heat-capacity})  we obtain the heat capacity as
\begin{equation}\label{heat-cap1a}
 C_+ \simeq\frac{2\pi r_+^2
}{3s^3}\Big[\epsilon \alpha (30 r_++7s-60r\ln2)-3s^3\Big] \, .
\end{equation}
Expression  (\ref{heat-cap1a}) implies that $C_+$ does not   diverge and
thus we do not have a second-order phase transition. In the left graph of
 Fig. \ref{Fig:3b} we depict   $C_+$ as a function of the horizon. As we can
see, in the  $\epsilon=0$ case we have  $C_+<0$ due to the negative derivative
of the temperature, as expected for the  the Reissner Nordstr\"om
 black hole.  Nevertheless, for $\epsilon>0$ we   obtain positive heat
capacity. This is one of the main results of the present work, namely that
  $f(T)$ modifications
improve the thermodynamic stability. Note that this is not the case in other
gravitational modifications, since for instance in $f(R)$ gravity the heat
capacity is  positive
only conditionally  \cite{Elizalde:2020icc,Nashed:2020kdb,Nashed:2020mnp}.

We close this subsection by the examination of the Gibb's free energy. In terms
of the the mass, temperature and entropy at
 the event horizon this is defined  as
\cite{Zheng:2018fyn,Kim:2012cma}:
\begin{equation}
\label{enr1}
G(r_+)={\cal{M}}(r_+)-T(r_+)S(r_+)\,.
\end{equation}
Inserting  (\ref{hor-mass-rad1a}), (\ref{m44-ee})
 and (\ref{ent1}) into (\ref{enr1}), we obtain
\begin{equation} \label{m77}
 {G_+}\equiv G(r_+) =\frac{3s^3(3s^2+r_+{}^2)+\epsilon\{\alpha[30r_+{}
^3ln(2r_+)+31sr_+{}^2+3s^3r_+{}^2-30s^2r_+-231s^3]\}+3\epsilon
s^3r_+(r_+c_2-c_1)}{12r_s{}^3}.
\end{equation}
In the right graph of Fig. \ref{Fig:3b} we depict the behavior of Gibb's free
energy.
As we observe it is always positive, for both  $\epsilon=0$ and
$\epsilon>0$.

\subsection{Thermodynamics of the black hole solution with $\mathbb{Q}_1(r)\neq
0$}\label{S5b}

\begin{figure}[ht]
\centering
 \includegraphics[scale=0.3]{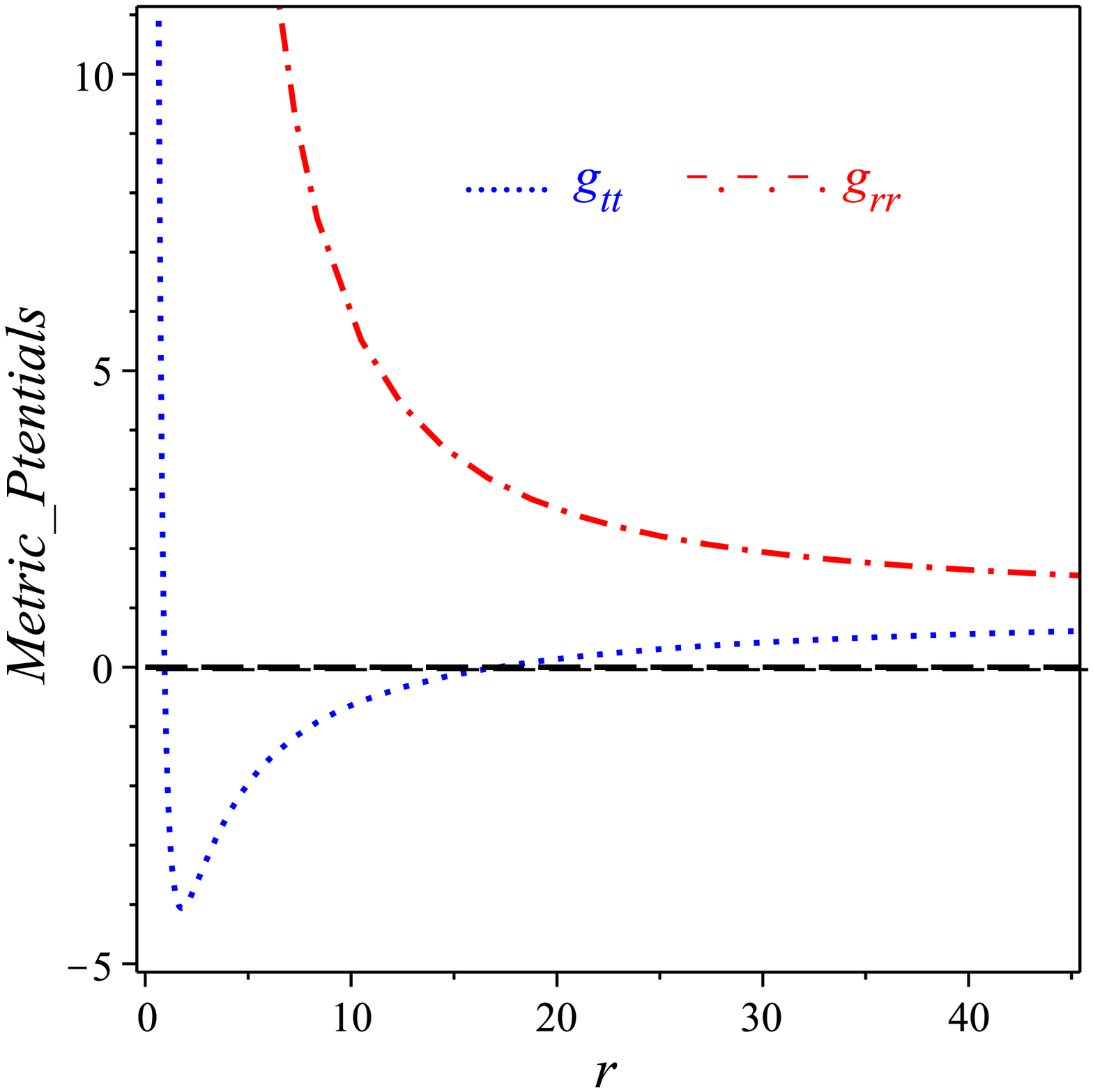}
 \includegraphics[scale=0.3]{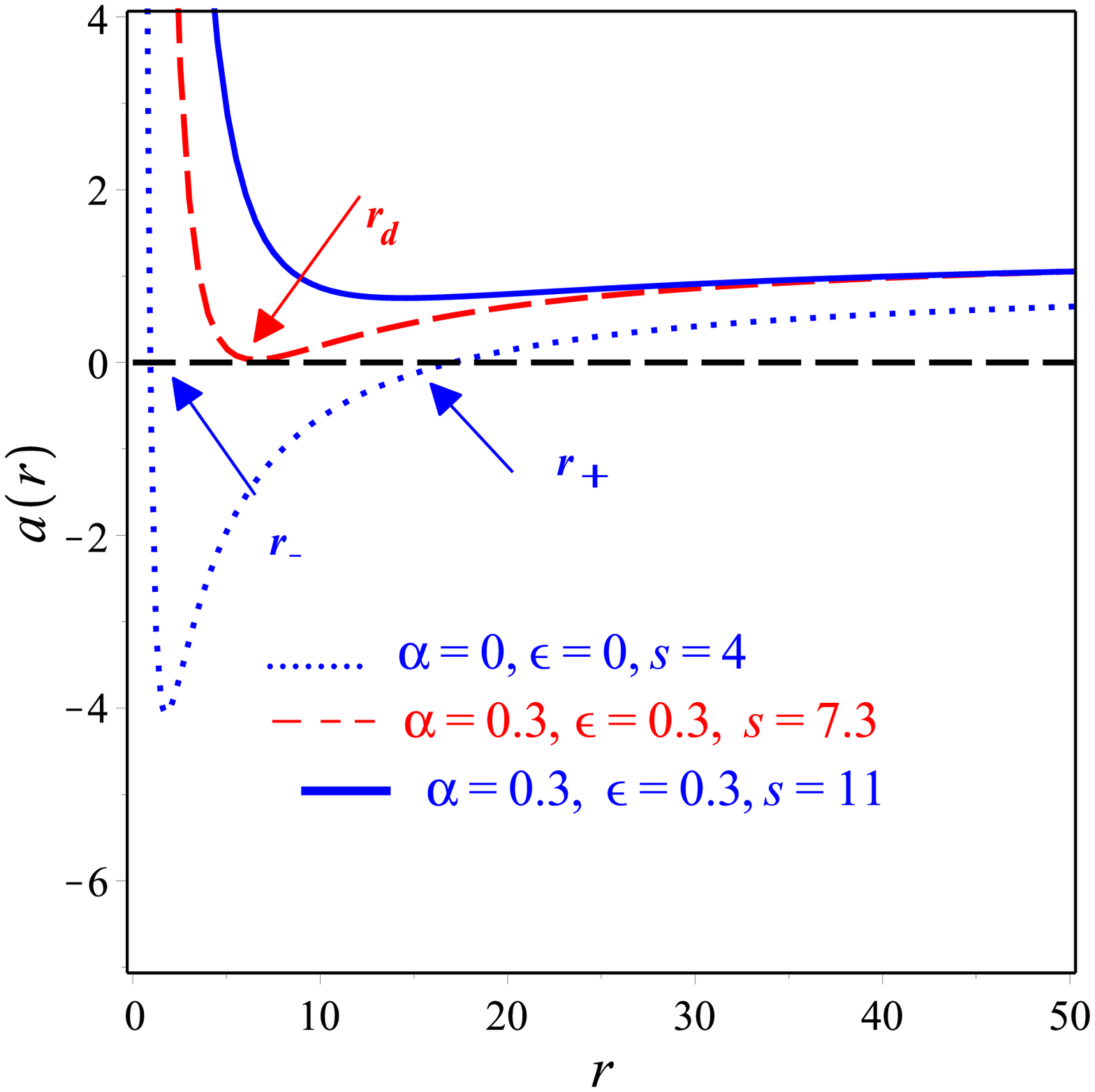}
\caption{{\it{Left graph:The two metric potentials $g_{tt}\equiv a(r)$ and
$g_{rr}\equiv b(r)$ given in
(\ref{mets2}),(\ref{mets2bb}) versus $r$, for the black hole solution with
$\mathbb{Q}_1(r)\neq
0$, for $M=9$, $c_1=c_2=1$, $\alpha=0.1$,
$\epsilon=0.1$,  and
$s=4$, in Planck
mass units. Right graph: The   metric potential  $a(r)$ versus $r$,
for $M=9$, $c_3=1$, $c_4=1$ and various values of  the model parameters
$\alpha$, $\epsilon$ and $s$,   in Planck
mass units.  $r_-$ and  $r_+$ are the inner and outer horizons
respectively, while $r_d $ is the degenerate horizon in which the above two
coincide.  }}}
\label{Fig:4}
\end{figure}

In this subsection we repeat the above thermodynamic analysis in the case of
the black hole solution for $\mathbb{Q}_1(r)\neq
0$ given in     (\ref{mets2}),(\ref{mets2bb}).
In the left graph of Fig. \ref{Fig:4}  we depict the metric
potentials $a(r)$ and $b(r)$, and as we observe $a(r)$ may exhibit two horizons
while $b(r)$ does not. In  the right graph of Fig. \ref{Fig:4} we present
$a(r)$. Similarly to the previous subsection,  we see that for small
$\alpha$ values, namely small deviations from general relativity,
 we   obtain  two horizons, however as $\alpha$ increases there is a specific
value in which the two horizons become degenerate ($r_{-}=r_{+}=r_d $), while
for larger values  the horizon disappears and the central singularity
becomes a naked one. However, the interesting feature is that for the
same $\alpha$ value, the parameter $s$ that quantifies the charge profile also
affects the horizon structure, and in particular   larger $s$ leads to the
appearance of the naked singularity.

\begin{figure}[ht]
\centering
 \includegraphics[scale=0.3]{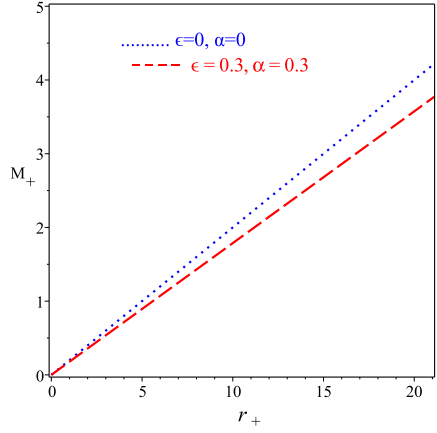}
\caption{{\it{ The mass-radius relation (\ref{hor-mass-rad1ac}) for the black
hole solution with
$\mathbb{Q}_1(r)\neq
0$, for   $c_3=1$, $c_4=0$, and $s=4$,
and various values of  the model parameters
$\alpha$, $\epsilon$ in Planck
mass units.  }}}
\label{Fig:4mass}
\end{figure}

The mass-radius relation  takes the   form
\begin{eqnarray} \label{hor-mass-rad1ac}
 {M_+}
=\frac{s^2+r_+{}^2}{2r_+}+\epsilon\left[\frac{3sc_3(2s^2+2r_+{}^2+sr_+{}^2)
+3s^2c_4r_++56\alpha (s^2+r_+{}^2)}{s^2r_+}\right]\,,
\end{eqnarray}
and it is
plotted   in Fig. \ref{Fig:4mass}, where we can verify that
  $M_+$ is  always   positive.

\begin{figure}[ht]
\centering
\includegraphics[scale=0.3]{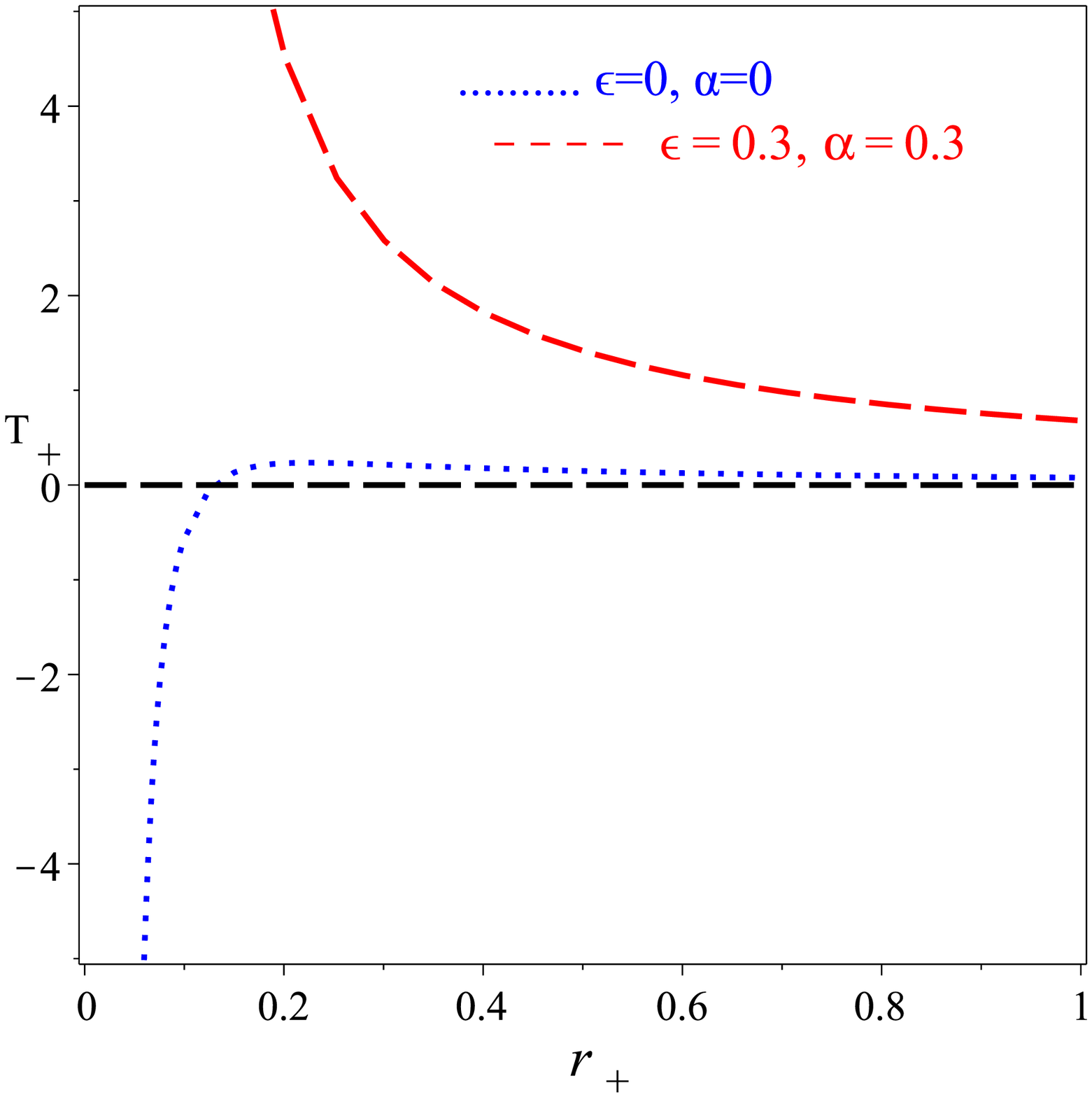}
 \includegraphics[scale=0.3]{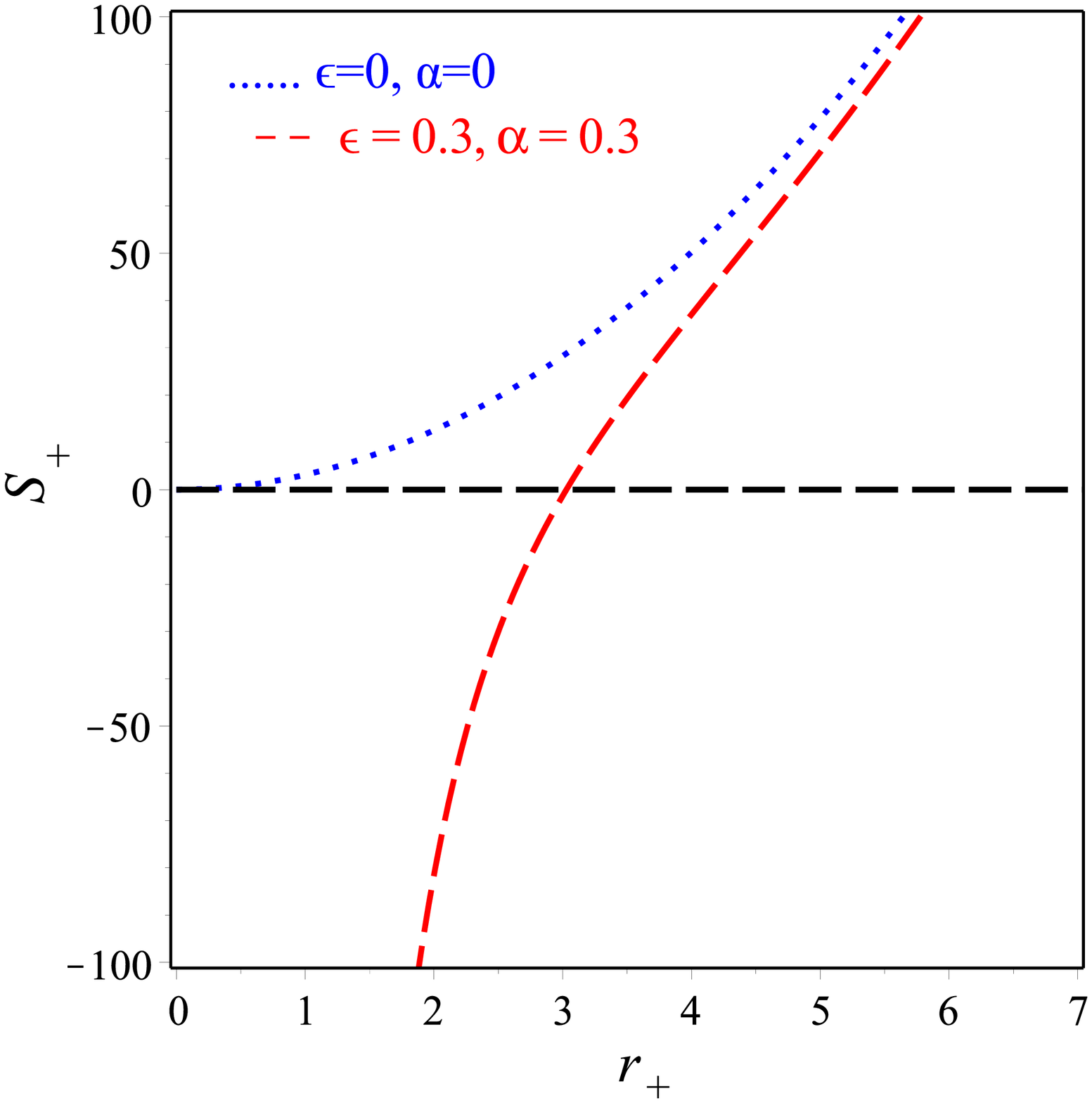}
\caption{{\it{ The temperature  (left graph) and entropy   (right
graph) versus the horizon,
 for the black hole solution with
$\mathbb{Q}_1(r)\neq
0$, for   $M=9$, $c_3=c_4=1$ and $s=4$,
and various values of  the model parameters
$\alpha$, $\epsilon$ in Planck
mass units.
 }}}
\label{Fig:5}
\end{figure}

For the temperature (\ref{tempdef}) we obtain
 \begin{eqnarray} \label{m44ec}
{T_+} \simeq\frac{3s^2(r_+{}^2-s^2)+\epsilon [r_+{}^2(56\alpha
+3s^2c_3+6c_3s)+6s^2(4\alpha-sc_3)]}{12\pi s^2r_+{}^3}\,.
\end{eqnarray}
 Moreover, for the entropy (\ref{ent}) we find
\begin{eqnarray} \label{ent1c}
&&{S_+}_  \simeq\frac{\pi}{r_+{}^3}\left[r_+{}^5-2\epsilon \alpha
M(r_+M-2s^2)
\right]\,,
\end{eqnarray}
which again for  $\epsilon=0$ recovers the general relativity result.
In   Fig. \ref{Fig:5} we depict the   temperature and  entropy versus the
horizon, for various values of the model parameters. We mention that in this
case both temperature and entropy may acquire negative values, however the
entropy, which is always quadratically increasing, is positive  when
$r_+>r_d$.

\begin{figure}[ht]
\centering
 \includegraphics[scale=0.33]{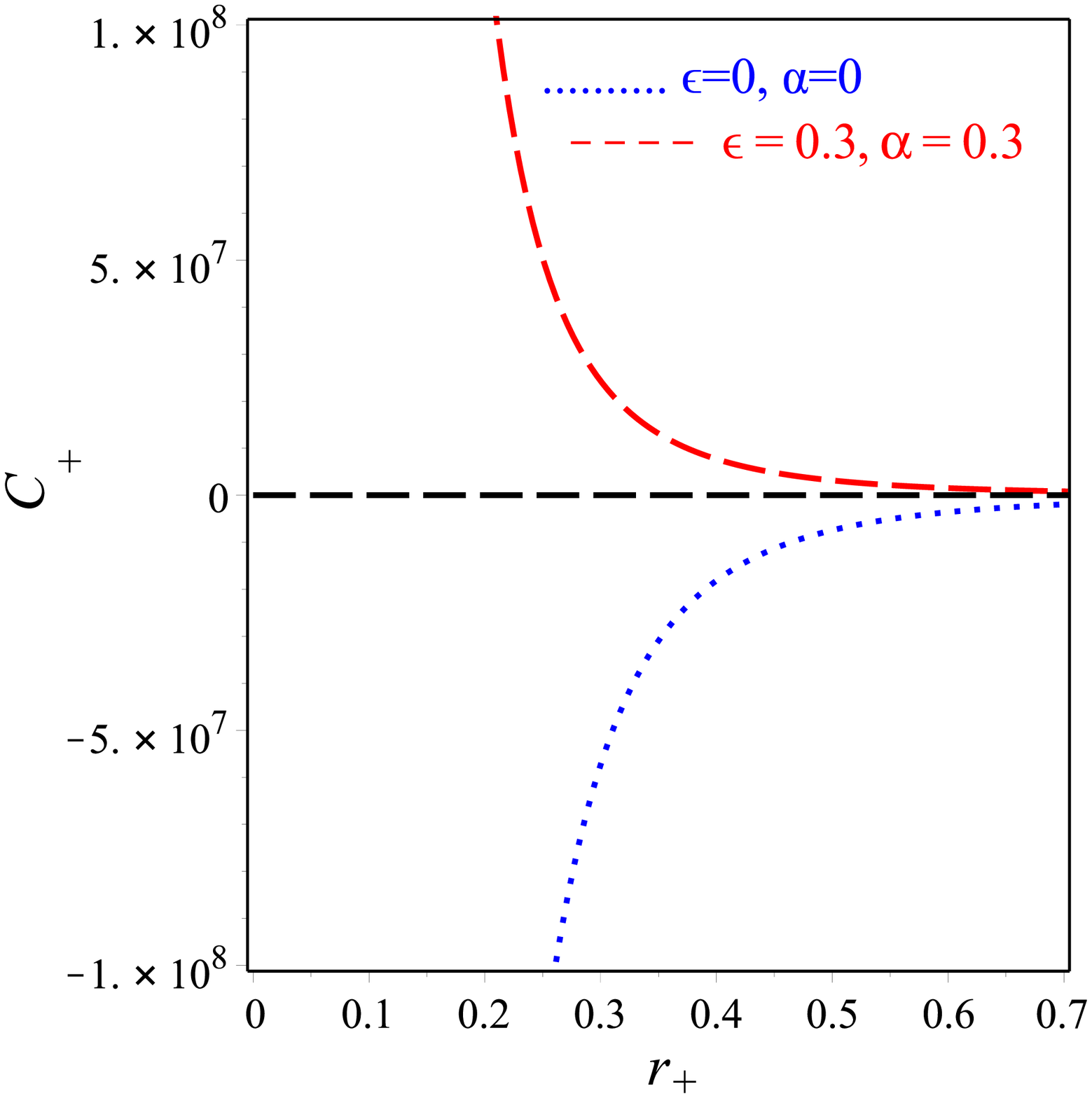}
 \includegraphics[scale=0.33]{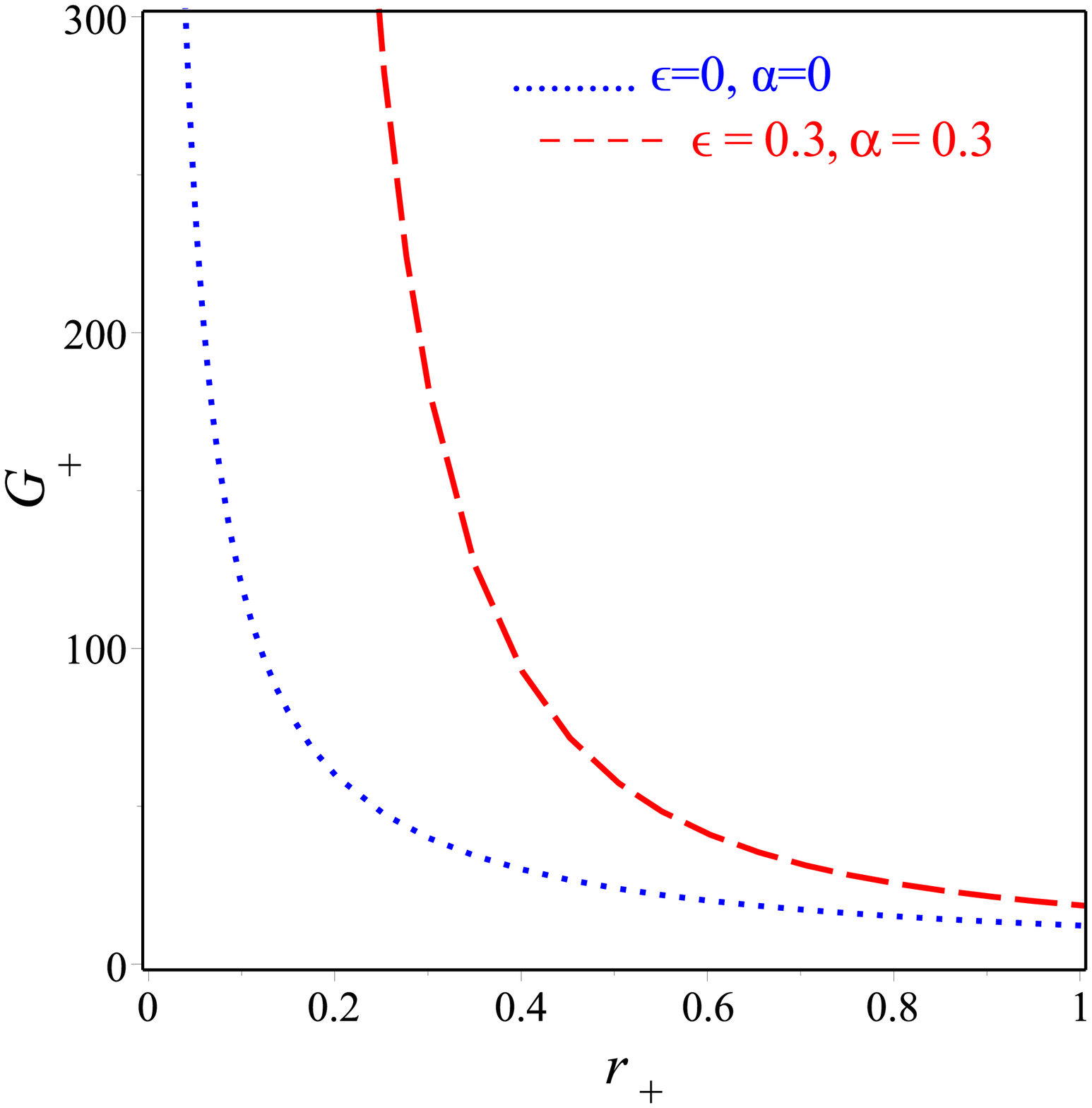}
\caption{{\it{
 The heat capacity  (left graph) and the Gibb's energy  (right
graph) versus the horizon,
 for the black hole solution with
$\mathbb{Q}_1(r)\neq
0$, for   $M=9$, $c_3=c_4=1$ and $s=4$,
and various values of  the model parameters
$\alpha$, $\epsilon$ in Planck
mass units.}}}
\label{Fig:5b}
\end{figure}

For the   heat capacity $
C_+  =\frac{\partial
M_+}{\partial r_+} \left(\frac{\partial T_+}{\partial r_+}\right)^{-1}
$, using
(\ref{hor-mass-rad1ac}) and (\ref{m44ec}), we acquire
\begin{equation}\label{heat-cap1ac}
{C_+}   \simeq-2\pi(2s^2+r_+{}^2-2s^2\epsilon c_3-8-\epsilon \alpha)
\, .
\end{equation}
Expression  (\ref{heat-cap1ac}) implies that $C_+$ does not   diverge and
therefore we do not have a second-order phase transition. In the left graph of
 Fig. \ref{Fig:5b} we present   $C_+$ as a function of the horizon. As we can
see, in the  $\epsilon=0$ case we have  $C_+<0$ due to the negative derivative
of the temperature, as expected for the  the Reissner Nordstr\"om
 black hole.  Nevertheless, for $\epsilon>0$, namely in the case where
the $f(T)$ correction is switched on, we may obtain positive heat
capacity.
 Finally, for the Gibb's free energy $G(r_+)=M(r_+)-T(r_+)S(r_+)\,$,
using (\ref{hor-mass-rad1ac}), (\ref{m44ec}) and (\ref{ent1c}), we find
 \begin{eqnarray} \label{m77c}
&&{G_+} \simeq\frac{3s^2(r_+{}^2+s^2)+\epsilon[3
c_3s(2r_+{}^2+sr_+{}^2+6s^2)+6s^2r_+c_4+8\alpha(7r_+{}^2+11s^2)]}{12s^2r_+}.
\end{eqnarray}
In the right graph of Fig. \ref{Fig:5b} we present  Gibb's free
energy as a function of the horizon.
As we can see    for both  $\epsilon=0$ and $\epsilon>0$ it is always positive.

\section{Conclusions and discussion}
\label{Conclusions}

 We investigated the stability of motion  and thermodynamics in the case 
of    spherically symmetric
solutions in $f(T)$ gravity  using the perturbative approach.  In particular, we
considered small deviations from teleparallel equivalent of general relativity
and we extracted charged black hole solutions for two charge profiles, namely
with or without a perturbative correction in the charge distribution.
Firstly, we examined their asymptotic behavior showing that for large distances
they become Minkowski. Then we extracted various torsional and curvature
invariants, which revealed the presence of the central singularity as expected.
Moreover, we calculated the energy and the mass of the solutions. As we showed,
all
results recover the general relativity ones in the case where the $f(T)$
deviation goes to zero.

As a next step we investigated the     stability of motion around    the 
obtained
black hole solutions, by extracting and studying the geodesic
deviation of a  test particle in their gravitational
field. Assuming a secular orbit, we extracted the corresponding  stability of 
motion 
condition in terms of the metric potentials. As we saw, in the case where the
perturbative correction to the charge profile is absent the solution is always
stable, however in the case where it is present we obtained unstable regimes in
the parameter space.

Additionally, we performed a detailed analysis of the thermodynamic properties
of the black hole solutions. In particular, we extracted the inner (Cauchy) and
outer (event) horizons, the mass profile, the temperature, the entropy, the
heat capacity and  the Gibb's free energy. As we showed,    for
small $\alpha$ values, namely small deviations from general relativity,
 we   obtain the two horizons, however as $\alpha$ increases there is a
specific value in which the two horizons become degenerate, and
for larger values  the horizon disappears and the central singularity
becomes a naked one, a known feature of torsional
gravity. Furthermore,  we saw that for the
same $\alpha$ value, the parameter $s$ that    quantifies the charge
profile also affects the horizon structure, and in
particular   larger $s$ leads to the appearance of the naked singularity.

Concerning the temperature and entropy, we showed that   although there are
regimes in which they become negative, for  $r_+>r_d$ they are always
positive definite. Concerning  the heat
capacity we saw that it does not   diverge and
thus we do not have a second-order phase transition. However, the most
interesting result is that  it becomes positive for larger
deviations from general relativity, which shows that $f(T)$ modifications
improve the thermodynamic stability, which is not the case in other
gravitational modifications.
Finally, for the Gibb's free energy, we showed that it is always positive, for
all torsional additions and for both charge-profile cases.

In summary, the present work indicates that torsional modification of gravity
may have an advantage comparing to other gravitational modification classes,
when stability issues are raised, which may serve as an additional motivation
for the corresponding investigations. One particular interesting issue is to
investigate in detail whether torsional modified gravity leads to smoother
(weaker) central singularities  comparing to general relativity or curvature
modified gravity. This issue will be the focus of interest of a separate
project.



\end{document}